\documentclass[11pt]{article}

\usepackage{bm,amsmath,amssymb,a4wide,graphicx,subfigure,wasysym,amsfonts,cite}

\usepackage[hyperindex,breaklinks]{hyperref}

\begin{document}

\title{\textbf{Magnetic fields in turbulent quark matter and magnetar bursts}}

\author{Maxim Dvornikov$^{a,b}$\thanks{maxdvo@izmiran.ru}
\\
$^{a}$\small{\ Pushkov Institute of Terrestrial Magnetism, Ionosphere} \\
\small{and Radiowave Propagation (IZMIRAN),} \\
\small{108840 Troitsk, Moscow, Russia;} \\
$^{b}$\small{\ Physics Faculty, National Research Tomsk State University,} \\
\small{36 Lenin Avenue, 634050 Tomsk, Russia}}

\date{}

\maketitle

\begin{abstract}
We analyze the magnetic field evolution in dense quark matter with
unbroken chiral symmetry, which can be found inside quark and hybrid stars. The magnetic field evolves owing to the chiral magnetic effect in the presence of the electroweak interaction between quarks. In our study, we also take into account the magnetohydrodynamic turbulence effects in dense quark matter. We derive the kinetic equations for the spectra of the magnetic helicity density and the magnetic energy density as well as for the chiral imbalances. On the basis of the numerical solution
of these equations, we find that turbulence effects are important for the behavior of small scale magnetic fields. It is revealed that, under
certain initial conditions, these magnetic fields behave similarly to the electromagnetic flashes of some magnetars. We suggest
that fluctuations of magnetic fields, described in frames of our model, which are created in the central regions of a magnetized compact star,
can initiate magnetar bursts.
\end{abstract}

\section{Introduction\label{sec:INTR}}

The emission of the strong electromagnetic radiation by some astrophysical
objects called anomalous X-ray pulsars (AXP) and soft $\gamma$-ray
repeaters (SGR) is likely to be associated with the transformation
of the magnetic energy in the star interiors to the energy of hard
photons~\cite{MerPonMel15}. The estimates of the strength of internal large scale magnetic
fields necessary to explain the observed flashes give the
value $B\apprge10^{15}\,\text{G}$, making these compact stars one
of the most powerful magnets in the universe, or magnetars~\cite{MerPonMel15}.
Despite the existence of the numerous models for the generation of
such strong magnetic fields based on classical magnetohydrodynamics (MHD),
this issue is still an open problem for the modern astrophysics. Note
that the actual mechanism for the transfer of the internal magnetic
energy to the stellar magnetosphere is unclear either. Some of the
popular models for the description of magnetars are reviewed in Ref.~\cite{TurZanWat15}.

Recently several attempts to explain the processes happening in magnetars
with help of the elementary particle physics methods were undertaken. For
example, the chiral magnetic effect (CME)~\cite{MirSho15}, which
leads to the magnetic field instability resulting in the enhancement
of a seed field, potentially can be used for the explanation of magnetic
fields in magnetars. Note that, the high linear velocities of pulsars~\cite{WanLaiHan06} can be also accounted for with help of CME. These astrophysical 
applications of CME are reviewed in Ref.~\cite{Kha15}.

Recently, in Ref.~\cite{DvoSem15b}, we proposed the new model for
the generation of strong magnetic fields in magnetars based on the
magnetic field instability driven by the parity violating electroweak
interaction between background fermions. In this model the contribution
of the electroweak interaction was treated as a correction to CME. Note that
the idea about the magnetic field instability induced by the parity violating
interaction was put forward earlier in Ref.~\cite{BoyRucSha15}.
%Despite the criticism in Refs.~\cite{SigLei16,KapRedSen16} of the original model in Ref.~\cite{DvoSem15b}, in Refs.~\cite{Dvo16b,Dvo16d,DvoSem17} we demonstrated that the main results of our approach are still valid.

Then, in Ref.~\cite{Dvo16a}, the same idea about the magnetic
field amplification driven by the electroweak interaction was implemented
in dense quark matter in a hybrid star (HS) or in a quark star (QS).
Sometimes QS is also called a strange star because of the presence
of the $s$ quark component. Although these compact stars have not
been observed yet, there is a quite strong theoretical motivation
for their existence~\cite{Gle00}.

In this work we continue the study of the magnetic field evolution
in dense quark matter where the chiral symmetry is restored and, thus, CME along with the electroweak correction can take place. Since the
typical time for the magnetic field growth found in Ref.~\cite{Dvo16a}
appeared to be much shorter than that in Ref.~\cite{DvoSem15b},
the magnetic field amplification can happen in earlier stages of the
stellar evolution when the effects of the turbulent motion of stellar matter are sizable. Thus
the main motivation of the present work will be the study of the magnetic
field evolution accounting for both CME, and the electroweak interaction between quarks, and the effects of the matter turbulence. Recently analogous problem in the early universe
plasma was considered in Ref.~\cite{DvoSem16}. The influence of the turbulence on the chiral MHD was also studied in Refs.~\cite{Vil17,Rog17}.

The present paper is organized as follows. In Sec.~\ref{sec:MODEL},
we derive the main equations describing the evolution of the magnetic
field in our model. These evolution equations are solved
numerically in Sec.~\ref{sec:NUMRES}. The results of the numerical
simulations are applied in Sec.~\ref{sec:APPL} to explain the electromagnetic
radiation of magnetars. In Sec.~\ref{sec:CONCL} we discuss our
results. The calculation of the drag time in dense quark matter is
given in Appendix~\ref{sec:TAUDCALC}.

\section{Model for the magnetic field evolution in dense quark matter\label{sec:MODEL}}

In this section, we shall establish the system of the evolution equations
for the spectra of the magnetic helicity density $h(k,t)$ and the
magnetic energy density $\rho_{\mathrm{B}}(k,t)$ as well as for the
chiral imbalances $\mu_{5(u,d)}$ of $u$ and $d$ quarks. This system
will take into account the effects of the turbulent matter motion.

First, let us briefly remind the quark matter properties inside a compact star. If matter density in the star interior is higher than the nuclear one $\sim 0.15 \,\text{fm}^{-3}$, quarks can lose their correlation with individual nucleons and quark matter is formed~\cite{Gle00}. The formation of quark matter is likely to take place in the vicinity of the stellar core. In this case, one deals with HS where the quark matter core is surrounded by the neutron crust. In the simplest situation, the QCD phase transition happens in the neutron matter, i.e. only the lightest $u$ and $d$
quarks are present. Accounting for the electric charge conservation, we get that $n_d = 2 n_u$, where $n_{u,d}$ are the densities of $u$ and $d$ quarks. We shall consider this relation between quark densities as one extreme situation which can happen in HS.

There is another possibility for the formation of quark matter in a compact star. It takes place if the following hypothesis is valid~\cite{Gle00}: the absolute ground state of the strong interactions is implemented in matter where $u$, $d$, and $s$ quarks are in approximately equal proportion, i.e. $n_u \approx n_d \approx n_s$. Here $n_s$ is the number density of $s$ quarks. A compact star, where quark matter with such properties is present, is called QS or a strange star. Note that QS can originate immediately after a 
supernova explosion without passing a neutron star stage. In principle, some models of nuclear interactions allow the existence of an admixture of $s$ quarks even in HS~\cite{Gle00}. In our work, we shall consider the situation when $n_u = n_d = n_s$ as another extreme case which is implemented in QS.

Let us consider a degenerate quark matter consisting of $u$ and $d$
quarks. As we mentioned above, this kind of matter can be found in the core of HS. We shall assume that the matter density is high enough for
the chiral symmetry to be restored. In this case CME can happen in this system. In Ref.~\cite{Dvo16a},
accounting for the electroweak interaction between quarks, we showed
that the anomalous electric current $\mathbf{J}_{5}$ along the external
magnetic field $\mathbf{B}$ is induced in this matter,
\begin{equation}\label{eq:Jind}
  \mathbf{J}_{5} = \Pi\mathbf{B},
  \quad
  \Pi = \frac{1}{2\pi^{2}}\sum_{q=u,d}e_{q}^{2}
  \left(
    \mu_{5q}+V_{5q}
  \right),
\end{equation}
where $e_{u}=2e/3$ and $e_{d}=-e/3$ are the electric charges of
quarks, $e>0$ is the elementary charge, $\mu_{5q}=\left(\mu_{q\mathrm{R}}-\mu_{q\mathrm{L}}\right)/2$
is the chiral imbalance, $\mu_{q\mathrm{R,L}}$ are the chemical potentials
of right and left quarks, $V_{5q}=\left(V_{q\mathrm{L}}-V_{q\mathrm{R}}\right)/2$,
and $V_{q\mathrm{L,R}}$ are the effective potentials of the electroweak
interaction of left and right quarks with background fermions.

The explicit expression for $V_{5q}$ in $ud$ quark matter was found in Ref.~\cite{Dvo15}. It is proportional to the number densities of quarks, $V_{5(u,d)} \sim G_{\mathrm{F}} n_{d,u}$, where $G_{\mathrm{F}}=1.17\times10^{-5}\,\text{GeV}^{-2}$ is the
Fermi constant. Here we take into account that the interactions of identical fermions, i.e. $uu$ and $dd$ interactions, do not contribute to $V_{5q}$, as found in Ref.~\cite{Dvo14}. If we suppose that the quark matter is electrically neutral, i.e. $n_d = 2 n_u$, one gets
that $V_{5u}=0.28G_{\mathrm{F}}n_{0}$ and $V_{5d}=0.18G_{\mathrm{F}}n_{0}$,
where $n_{0}=n_{u}+n_{d}$ is the total number density of
quarks.
Taking that $n_{0}=1.8\times10^{38}\thinspace\text{cm}^{-3}$,
which corresponds to a typical quark matter in HS, we obtain that
$V_{5u}=4.5\thinspace\text{eV}$ and $V_{5d}=2.9\thinspace\text{eV}$~\cite{Dvo16a}.

We shall describe the evolution of the magnetic field in an isotropic
matter using the magnetic helicity density $h$ and the magnetic energy
density $\rho_{\mathrm{B}}$, which are defined as
\begin{equation}
  h(t) = \frac{1}{V}
  \int\mathrm{d}^{3}x
  \left(
    \mathbf{A}\cdot\mathbf{B}
  \right),
  \quad
  \rho_{\mathrm{B}}(t)=\frac{1}{2V}\int\mathrm{d}^{3}xB^{2},
\end{equation}
where $\mathbf{A}$ is the vector potential of the magnetic field
and $V$ is the normalization volume. It is convenient to express
$h(t)$ and $\rho_{\mathrm{B}}(t)$ in terms of their spectral representations
$h(k,t)$ and $\rho_{\mathrm{B}}(k,t)$ in the form,
\begin{equation}
  h(t) = \int\mathrm{d}kh(k,t),
  \quad
  \rho_{\mathrm{B}}(t) = \int\mathrm{d}k\rho_{\mathrm{B}}(k,t),
\end{equation}
where the integration is over all the range of the wave number $k$
variation.

In Ref.~\cite{DvoSem16}, we studied the evolution of the magnetic
field in the relativistic electron fluid accounting for CME and the MHD turbulence. The effects of the turbulence
were simulated by the replacement of the fluid velocity $\mathbf{v}$,
obeying the Navier-Stokes equation, with the Lorentz force~\cite{Sig02}:
\begin{equation}\label{eq:vFL}
  \mathbf{v} = \frac{\tau_{\mathrm{D}}}{\rho_{\mathrm{E}}+p}
  \left(
    \mathbf{J}\times\mathbf{B}
  \right),
\end{equation}
where $\rho_{\mathrm{E}}$ is the energy density of the fluid, $p$
is the pressure, $\mathbf{J}$ is the total electric current, and
$\tau_{\mathrm{D}}$ is the drag time, which is a phenomenological
parameter meaning the typical Coulomb collision time in plasma~\cite{Cor97}.
The computation of $\tau_{\mathrm{D}}$ in the degenerate quark matter
is given in Appendix~\ref{sec:TAUDCALC} (see Eq.~(\ref{eq:tauDfin})
for the final result). If we study the degenerate relativistic quark
gas, then $p=\rho_{\mathrm{E}}/3$, $\rho_{\mathrm{E}}=\left(\mu_{u}^{4}+\mu_{d}^{4}\right)/4\pi^{2}=2\times10^{-2}\mu_{0}^{4}$,
$\mu_{q}=\left(3\pi^{2}n_{q}\right)^{1/3}$ is the mean chemical potential
of each quark component, and $\mu_{0}=\left(3\pi^{2}n_{0}\right)^{1/3}=346\,\text{MeV}$.

In general situation, other terms in the Navier-Stokes equation can result in the turbulent motion of matter. By using the approximation in Eq.~\eqref{eq:vFL}, we restrict ourselves to the MHD turbulence, i.e. we assume that the Lorentz force is dominant in the Navier-Stokes equation. The MHD turbulence was found in Ref.~\cite{Cam07} to take place when both the Reynolds number, $\text{Re} = vl / \nu$, and the magnetic Reynolds number, $\text{Re}_\mathrm{B} = vl\sigma_\mathrm{cond}$, are large. Here $v$ and $l$ are the typical velocity and length scales of the matter motion, $\nu$ is the kinematic viscosity, and $\sigma_\mathrm{cond}$ is the quark matter conductivity. The Reynolds numbers in HS/QS were estimated in Ref.~\cite{XuBus01} as $\text{Re} \sim 10^8$ and $\text{Re}_\mathrm{B} \sim 10^{16}$. Thus we can see that the MHD turbulence should be dominant since $\text{Re} \gg 1$ and $\text{Re}_\mathrm{B} \gg 1$.

%The validity of the approximation in Eq.~\eqref{eq:vFL} will be discussed later in Sec.~\ref{sec:CONCL} on the basis of the results of numerical simulations made in Sec.~\ref{sec:NUMRES}.

Besides the requirement of the large Reynolds numbers, for the validity of Eq.~\eqref{eq:vFL} we should demand that the microscopic motion of a quark in the strong magnetic field does not affect the macroscopic motion of plasma under the influence of the Lorenz force $\mathbf{F}_\mathrm{L} \sim (\mathbf{J}\times\mathbf{B})$. This condition is fulfilled if $l_\mathrm{B} \ll \tau_\mathrm{D}$, where $l_\mathrm{B} \sim \mu_q / e_q B$ is the Larmor radius in a relativistic degenerate quark matter. Using Eq.~\eqref{eq:tauDfin}, one can see that the above inequality is satisfied for the lowest temperature $T_0 = 10^8\,\text{K}$ and the weakest magnetic field $B_0 = 10^{12}\,\text{G}$, which will be used in Secs.~\ref{sec:NUMRES} and~\ref{sec:APPL}.

In case of QS, the abundance of $s$ quarks, $Y_s = n_s / n_0$, depends on the chosen low energy approximation of strong interactions. Typically this abundance does not exceed $1/3$~\cite{MenProMel06}. We mentioned above that we can take that $n_u = n_d = n_s = n_0/3$ in QS as a limiting case. This relation between the number densities also follows from both the beta equilibrium and the matter electroneutrality~\cite{HaePotYak07}, and hence $\mu_u = \mu_d = 0.69 \mu_0$.

To find the contribution of the $s$ quark component to the anomalous current in Eq.~\eqref{eq:Jind} we first mention that an $s$ quark is quite heavy for a chiral symmetry to be restored for this quark component of matter~\cite{MenProMel06}. Therefore, there is no direct contribution  of $s$ quarks to the anomalous electric current in Eq.~\eqref{eq:Jind}. Instead, these quarks form a background which $u$ and $d$ quarks electroweakly interact with. Thus, the one can obtain that $V_{5(u,d)} = G_\mathrm{F} \left( \kappa_{u,d}' n_{d,u} + \kappa_{u,d}'' n_{s} \right)$, where $\kappa_{u,d}'$ and $\kappa_{u,d}''$ are the constant coefficients. The explicit form of these coefficients can be found analogously to Ref.~\cite{Dvo15}. Then, accounting for the chosen number densities of quarks in QS, $n_q = n_0/3$, we get that $V_{5u} = 1.1\,\text{eV}$ and $V_{5d} = 4.1\,\text{eV}$.

Generalizing the results of Ref.~\cite{DvoSem16} by considering
the chiral imbalances of $u$ and $d$ quarks and including the nonzero
effective potentials $V_{5q}$, we get the following system of the
evolution equations for $h(k,t)$, $\rho_{\mathrm{B}}(k,t)$, and $\mu_{5q}$:
\begin{align}%\label{eq:syskindm}
  \frac{\partial h(k,t)}{\partial t} = &
  -2k^{2} \eta_\mathrm{eff} h(k,t) + 4 \alpha_{-} \rho_{\mathrm{B}}(k,t),
  \label{eq:dhdt}
  %\nonumber
  \displaybreak[1]
  \\
  \frac{\partial\rho_{\mathrm{B}}(k,t)}{\partial t} = & 
  -2k^{2} \eta_\mathrm{eff} \rho_{\mathrm{B}}(k,t) + \alpha_{+} k^{2}h(k,t),
  \label{eq:drhodt}
  %\nonumber
  \displaybreak[1]
  \\
  \frac{\mathrm{d}\mu_{5u}(t)}{\mathrm{d}t} = &   
  - \frac{\pi\alpha_{\mathrm{em}}}
  {2\mu_{u}^{2}}
  \frac{4}{9}
  \int\mathrm{d}k
  \frac{\partial h(k,t)}{\partial t} -
  \Gamma_{u}\mu_{5u}(t),
  \label{eq:dmuudt}
  %\nonumber
  \displaybreak[1]
  \\
  \frac{\mathrm{d}\mu_{5d}(t)}{\mathrm{d}t} = &   
  - \frac{\pi\alpha_{\mathrm{em}}}
  {2\mu_{d}^{2}}
  \frac{1}{9}
  \int\mathrm{d}k
  \frac{\partial h(k,t)}{\partial t} -
  \Gamma_{d}\mu_{5d}(t).
  \label{eq:dmuddt}
\end{align}
where, analogously to Ref.~\cite{DvoSem16}, we introduce the effective magnetic diffusion coefficient $\eta_\mathrm{eff}$ and the effective dynamo parameters $\alpha_{\pm}$, which account for the MHD turbulence, as
\begin{align}\label{eq:etaalpha}
  \eta_\mathrm{eff} = &
  \frac{F_{Q}^{5/6}}{\sigma_{\mathrm{cond}}} +
  \frac{4}{3}\frac{\tau_{\mathrm{D}}}{\rho_{\mathrm{E}}+p}
  \int\mathrm{d}k'\rho_{\mathrm{B}}(k',t),
  \notag
  \\
  \alpha_{\pm} = &
  \frac{2\alpha_{\mathrm{em}}F_{Q}^{5/6}}{\pi\sigma_{\mathrm{cond}}}
  \left\{
  \frac{4}{9}
    \left[
      \mu_{5u}(t)+V_{5u}
    \right] +
    \frac{1}{9}
    \left[
      \mu_{5d}(t)+V_{5d}
    \right]
  \right\}
  \notag
  \displaybreak[1]
  \\
  & \mp
  \frac{2}{3}\frac{\tau_{\mathrm{D}}}{\rho_{\mathrm{E}}+p}
  \int\mathrm{d}k' k^{\prime2}h(k',t).
\end{align}
In Eqs.~\eqref{eq:dmuudt} and~\eqref{eq:dmuddt}, we take into account the helicity flip of quarks in their collisions
by introducing the coefficients $\Gamma_{q}=e_{q}^{6}\mu_{q}/8\pi^{5}$~\cite{Dvo16a}. If we study the case of QS, $\Gamma_{q}$ will have the same value since the probability for $u$ and $d$ quarks to spin-flip in their scattering off unpolarized $s$ quarks is much less than that in the mutual scattering of $u$ and $d$ quarks~\cite{Dvo16a}.

The derivative $\partial h / \partial t$ in the integrands in Eqs.~\eqref{eq:dmuudt} and~\eqref{eq:dmuddt} should be substituted from Eq.~\eqref{eq:dhdt}. Note that, in this case, there is no direct contribution of the MHD turbulence in the evolution of $\mu_{5q}$, as found in Ref.~\cite{DvoSem16}.

The quark matter conductivity in Eq.~(\ref{eq:etaalpha}) as a function
of the matter temperature $T$ has the form~\cite{HeiPet93},
\begin{equation}\label{eq:sigmaT}
  \sigma_{\mathrm{cond}} = \sigma_{0}\frac{T_{0}^{5/3}}{T^{5/3}},
  \quad
  \sigma_{0}=3.15\times10^{22}\,\text{s}^{-1},
\end{equation}
where $T_{0}\sim(10^{8}-10^{9})\,\text{K}$ is the typical initial
temperature of a thermally relaxed HS/QS. To account for the energy
conservation law, we insert the quenching factor,
\begin{equation}\label{eq:Fqdef}
  F_{Q}=1-\frac{B^{2}}{(\mu_{u}^{2}+\mu_{d}^{2})T_{0}^{2}},
\end{equation}
in Eq.~(\ref{eq:etaalpha}); cf. Ref.~\cite{Dvo16a} and Eq.~(\ref{eq:sigmaT}).
The quenching factor in Eq.~(\ref{eq:Fqdef}) takes into account the anticorrelation between 
the temperature and the magnetic field~\cite{Dvo16b,Dvo16d}: $T^{2}=T_{0}^{2}F_{Q}$. The quenching factors in the kinematic $\alpha$-dynamo models analogous to that in Eq.~\eqref{eq:Fqdef} were discussed in Ref.~\cite{BraSub05}.

In our analysis, we assume that magnetic field in HS/QS evolves in homogeneous quark matter. This assumption implies that the chiral imbalances $\mu_{5q}$ in Eqs.~\eqref{eq:dmuudt} and~\eqref{eq:dmuddt} depend on time rather than on spatial coordinates (or equivalently on the wave number $k$). Hence we do not study, e.g., boundary effects on the magnetic field evolution. The generalization of Eqs.~\eqref{eq:dmuudt} and~\eqref{eq:dmuddt} taking into account the chiral imbalances  dependence on spatial coordinates was recently made in Ref.~\cite{BoyFroRuc15}.

Eqs.~\eqref{eq:dhdt}-\eqref{eq:dmuddt} should be completed
with the initial condition. We shall suppose that the initial spectrum
of the magnetic energy density has the Kolmogorov form~\cite{DvoSem15a},
\begin{equation}\label{eq:Kolspec}
  \rho_{\mathrm{B}}(k,0) = \mathcal{C}k^{-5/3},
  \quad
  \mathcal{C}=\frac{B_{0}^{2}}
  {3\left(k_{\mathrm{min}}^{-2/3}-k_{\mathrm{max}}^{-2/3}\right)},
\end{equation}
where $k_{\mathrm{min}}$ and $k_{\mathrm{max}}$ are the minimal
and maximal wave numbers and $B_{0}$ is the initial magnetic field.
For the definiteness we shall take that $k_{\mathrm{max}}=10k_{\mathrm{min}}$, where $k_{\mathrm{min}}$ will be the free parameter in our model. The
initial spectrum of the magnetic field is $h(k,0)=2\rho_{\mathrm{B}}(k,0)/k$,
which corresponds to the maximal initial helicity.

A seed magnetic field is supposed to emerge at the stages of a hot HS/QS when $T \gg T_0$. We assume that the seed field $B_0 = 10^{12}\,\text{G}$ is created owing to a forward cascade in hot matter of a compact star, i.e. when large scale eddies are converted into small scale ones~\cite{Dav15}.
%In Sec.~\ref{sec:NUMRES} we shall see that that maximal scale relevant for astrophysical applications is in the range $(10 - 10^3)\,\text{cm}$.
We do not discuss here the actual scenario for the seed magnetic field generation. However, it can be based on the classical MHD mechanisms without involving elementary particle physics approaches like CME. Indeed, the chiral phase transition in quark matter does not occur at the time of the compact star evolution, when the matter density is not high enough. Thus, $\mathbf{J}_5 = 0$ in Eq.~\eqref{eq:Jind} when $T \gg T_0$. The Kolmogorov spectrum for the energy density in Eq.~\eqref{eq:Kolspec} is known to be a good approximation~\cite{Dav15} when the seed magnetic field is created due to the forward cascade. When quark matter cools down to $T \sim T_0$, the chiral phase transition takes place and CME is present. Then, if $T \leq T_0$, the growth of a seed magnetic field is driven by CME under the influence of the electroweak interaction between quarks.

The initial chiral
imbalances are chosen as $\mu_{5u}(0)=\mu_{5d}(0)=1\,\text{MeV}$~\cite{Dvo16a}. This value of $\mu_{5(u,d)}(0)$ can be justified by the fact that the initial chiral imbalance should be created in a protostar at the moment just before the chiral phase transition. The process of the creation of $\mu_{5(u,d)}(0)$ is likely to be electroweak with the energy scale $\sim (m_d - m_u) \sim \mathcal{O}(\text{MeV})$, where $m_{u,d}$ are the vacuum masses of $u$ and $d$ quarks. It defines the initial value of $\mu_{5(u,d)}$. Moreover, in Refs.~\cite{DvoSem15a,DvoSem15b}, we analyzed the influence of initial values of the chiral imbalance on the magnetic field generated in our model. It turns out that, if $|\mu_{5q}(0)| \ll \mu_q$ (i.e., negative values of initial chiral imbalances can be also considered), the evolution of the magnetic field practically does not depend on $\mu_{5q}(0)$ since such initial chiral imbalance is washed out rapidly owing to the huge spin flip rate in fermion collisions in the compact star matter.

A successful amplification of a seed magnetic field requires equal signs of $V_{5q}$ and $h(k,0)$ since the growth of the magnetic field is driven by the electroweak interaction between quarks. We have shown above that $V_{5q}>0$. Hence the positive sign of $h(k,0)$ is necessary. If we studied the evolution of a large scale magnetic field in a compact star, as in Ref.~\cite{Dvo16a}, it would mean that the magnetic filed can be enhanced only in one hemisphere of a star since a seed field should have opposite helicities in differerent hemispheres. In the present work, we shall be mainly interested in the evolution of small scale magnetic fields with $1/k \sim (1-10^3)\,\text{cm}$, that is much less than the stellar radius (see Sec.~\ref{sec:NUMRES}). For such magnetic field, one can always find a domain with a proper sign of the initial helicity.

%The nonzero initial chiral imbalances appear due to the fact that left chiral fermions participate in direct Urca processes with the emission of left neutrinos which easily escape a compact star. Thus the number densities of left fermions should be less than that of right particles, resulting in $\mu_{5q}(0) = [\mu_{q\mathrm{R}}(0)-\mu_{q\mathrm{L}}(0)]/2 > 0$. The direct calculation of the electroweak effective potentials, made above, gives that $V_{5q} = (V_{q\mathrm{L}}-\mu_{q\mathrm{R}})/2 > 0$. The equal signs of $\mu_{5q}(0)$ and $V_{5q}$ allow one to have the amplification of a seed field with a positive initial helicity, which will be described in Sec.~\ref{sec:NUMRES}. Thus, the signs of $\mu_{5q}(0)$, $V_{5q}$, and $h(k,0)$, chosen in our work, are physically motivated.

%The growth of the magnetic field with negative initial helicity is also possible if both $\mu_{5q}(0)$ and $V_{5q}$ are negative. If the signs of $\mu_{5q}(0)$ and $V_{5q}$ are different, there are situations... 

If we introduce the dimensionless quantities,
\begin{align}
  \mathcal{H}(\kappa,\tau) & =
  \frac{\alpha_{\mathrm{em}}^{2}}{2\mu_{0}^{2}}h(k,t),
  \quad
  \mathcal{R}(\kappa,\tau) =
  \frac{\alpha_{\mathrm{em}}^{2}}{\mu_{0}^{2}k_{\mathrm{min}}}
  \rho_{\mathrm{B}}(k,t),
  \quad
  \mathcal{M}_{u,d}(\tau) =
  \frac{\alpha_{\mathrm{em}}}{\pi k_{\mathrm{min}}}
  \mu_{5(u,d)}(t),
  \nonumber
  \\
  \kappa = & \frac{k}{k_{\mathrm{min}}},
  \quad
  \tau = \frac{2k_{\mathrm{min}}^{2}}{\sigma_{0}}t,
  \quad
  \mathcal{V}_{u,d} =
  \frac{\alpha_{\mathrm{em}}}{\pi k_{\mathrm{min}}}V_{5(u,d)},
  \quad
  \mathcal{G}_{u,d} =
  \frac{\sigma_{0}\Gamma_{u,d}}{2k_{\mathrm{min}}^{2}},
\end{align}
where $\alpha_{\mathrm{em}}=e^{2}/4\pi=7.3\times10^{-3}$ is the fine
structure constant, then Eqs.~\eqref{eq:dhdt}-\eqref{eq:dmuddt}
can be rewritten in the form,
\begin{align}%\label{eq:syskindmls}
  \frac{\partial\mathcal{H}(\kappa,\tau)}{\partial\tau} = &
  -\kappa^{2}
  \left\{   
    F_{Q}^{5/6}+F_{Q}^{-1}K_{d}
    \int_{1}^{\kappa_{\mathrm{max}}}
    \mathrm{d}\kappa'\mathcal{R}(\kappa',\tau)
  \right\}
  \mathcal{H}(\kappa,\tau)
  \nonumber
  \displaybreak[1]
  \\
  & +
  \bigg\{
    0.22F_{Q}^{5/6}
    \left(
      4
      \left[
        \mathcal{M}_{u}(\tau)+\mathcal{V}_{u}
      \right] +
      \mathcal{M}_{d}(\tau) + \mathcal{V}_{d}
    \right)
    \nonumber
    \displaybreak[1]
    \\
    & +
    F_{Q}^{-1}K_{d}
    \int_{1}^{\kappa_{\mathrm{max}}}
    \mathrm{d}\kappa'\kappa^{\prime2}\mathcal{H}(\kappa',\tau)
  \bigg\}
  \mathcal{R}(\kappa,\tau),
  \label{eq:syskindmlsH}
  \displaybreak[1]
  \\
  \frac{\partial\mathcal{R}(\kappa,\tau)}{\partial\tau} = &
  -\kappa^{2}
  \left\{   
    F_{Q}^{5/6} + K_{d}F_{Q}^{-1}
    \int_{1}^{\kappa_{\mathrm{max}}}
    \mathrm{d}\kappa'\mathcal{R}(\kappa',\tau)
  \right\}
  \mathcal{R}(\kappa,\tau)
  \nonumber
  \displaybreak[1]
  \\
  & +
  \kappa^{2}
  \bigg\{
    0.22F_{Q}^{5/6}
    \left(
      4
      \left[
        \mathcal{M}_{u}(\tau)+\mathcal{V}_{u}
      \right] +
      \mathcal{M}_{d}(\tau) + \mathcal{V}_{d}
    \right)
    \nonumber
    \displaybreak[1]
    \\
    & -  
    F_{Q}^{-1}K_{d}
    \int_{1}^{\kappa_{\mathrm{max}}}
    \mathrm{d}\kappa'\kappa^{\prime2}\mathcal{H}(\kappa',\tau)
  \bigg\}
  \mathcal{H}(\kappa,\tau),
  \label{eq:syskindmlsR}
  \displaybreak[1]
  \\
  \frac{\mathrm{d}\mathcal{M}_{u}(\tau)}{\mathrm{d}\tau}= &   
  0.92F_{Q}^{5/6}
  \int_{1}^{\kappa_{\mathrm{max}}}
  \mathrm{d}\kappa
  \big[
    \kappa^{2}\mathcal{H}(\kappa,\tau)
    \nonumber
    \displaybreak[1]
    \\
    & -
    0.22
    \left(
      4
      \left[
        \mathcal{M}_{u}(\tau)+\mathcal{V}_{u}
      \right] +
      \mathcal{M}_{d}(\tau)+\mathcal{V}_{d}
    \right)
    \mathcal{R}(\kappa,\tau)
  \big] -
  \mathcal{G}_{u}\mathcal{M}_{u}(\tau),
  \label{eq:syskindmlsMu}
  \displaybreak[1]
  \\
  \frac{\mathrm{d}\mathcal{M}_{d}(\tau)}{\mathrm{d}\tau} = & 
  0.15F_{Q}^{5/6}
  \int_{1}^{\kappa_{\mathrm{max}}}
  \mathrm{d}\kappa
  \big[
    \kappa^{2}\mathcal{H}(\kappa,\tau)
    \nonumber
    \displaybreak[1]
    \\
    & -
    0.22
    \left(
      4
      \left[
        \mathcal{M}_{u}(\tau)+\mathcal{V}_{u}
      \right] +
      \mathcal{M}_{d}(\tau)+\mathcal{V}_{d}
    \right)
    \mathcal{R}(\kappa,\tau)
  \big] -
  \mathcal{G}_{d}\mathcal{M}_{d}(\tau),
  \label{eq:syskindmlsMd}
\end{align}
where $\kappa_{\mathrm{max}}=k_{\mathrm{max}}/k_{\mathrm{min}}$.
The quenching factor $F_{Q}$ in Eqs.~(\ref{eq:syskindmlsH})-\eqref{eq:syskindmlsMd} takes
the form,
\begin{equation}\label{eq:Fq}
  F_{Q} = 1-
  \frac{1.61k_{\mathrm{min}}^{2}}{\alpha_{\mathrm{em}}^{2}T_{0}^{2}}
  \int_{1}^{\kappa_{\mathrm{max}}}
  \mathrm{d}\kappa\mathcal{R}(\kappa,\tau).
\end{equation}
The coefficient in the turbulence terms in Eqs.~(\ref{eq:syskindmlsH})-\eqref{eq:syskindmlsMd}
reads,
\begin{equation}\label{eq:Kd}
  K_{d}=1.25\times10^{2}
  \frac{\sigma_{0}k_{\mathrm{min}}^{2}}
  {\alpha_{\mathrm{em}}^{3}T_{0}^{2}\mu_{0}}.
\end{equation}
Note that we also take into account the anticorrelation between 
the temperature and the magnetic field in the turbulence terms by introducing the factor
$F_{Q}^{-1}$ there.

If we study QS, one should multiply the pre-integral factor in Eq.~\eqref{eq:Fq} by 1.29,  the pre-integral factor in Eq.~\eqref{eq:syskindmlsMd} by 1.59, as well as take into account the additional factor 2.11 in Eq.~\eqref{eq:Kd}. Thus, one can see that $s$ quarks, considered in equal proportions with $u$ and $d$ quarks, amplify the effect of the turbulent motion of matter on the generation of magnetic fields by more than 2 times.

\section{Numerical solution of the evolution equations\label{sec:NUMRES}}

In this section, we present the numerical solution of the system in
Eqs.~(\ref{eq:syskindmlsH})-\eqref{eq:syskindmlsMd} with the initial conditions chosen in
Sec.~\ref{sec:MODEL}.

\begin{figure}
  \centering
  \subfigure[]
  {\label{1a}
  \includegraphics[scale=.11]{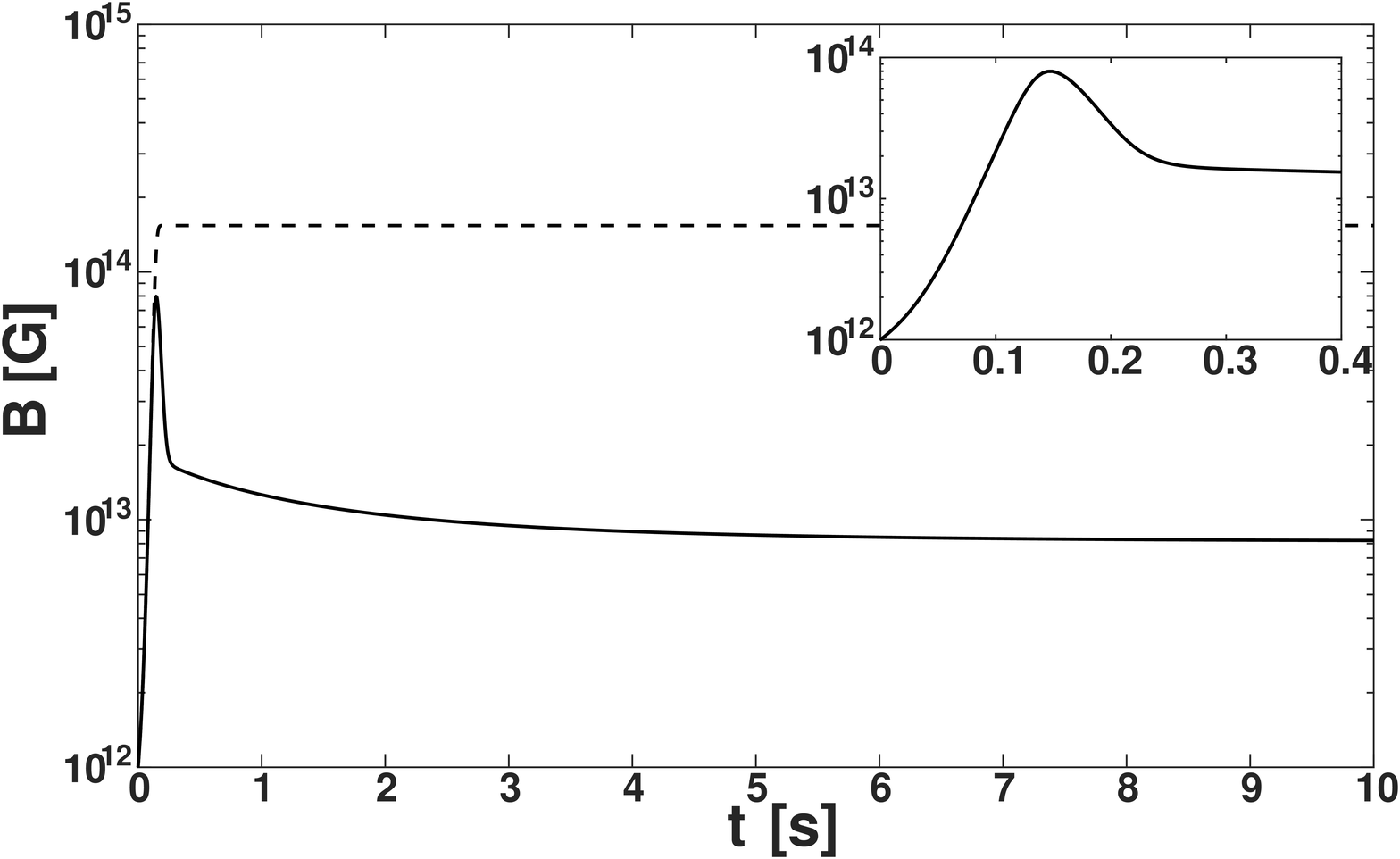}}
  \hskip-.7cm
  \subfigure[]
  {\label{1b}
  \includegraphics[scale=.11]{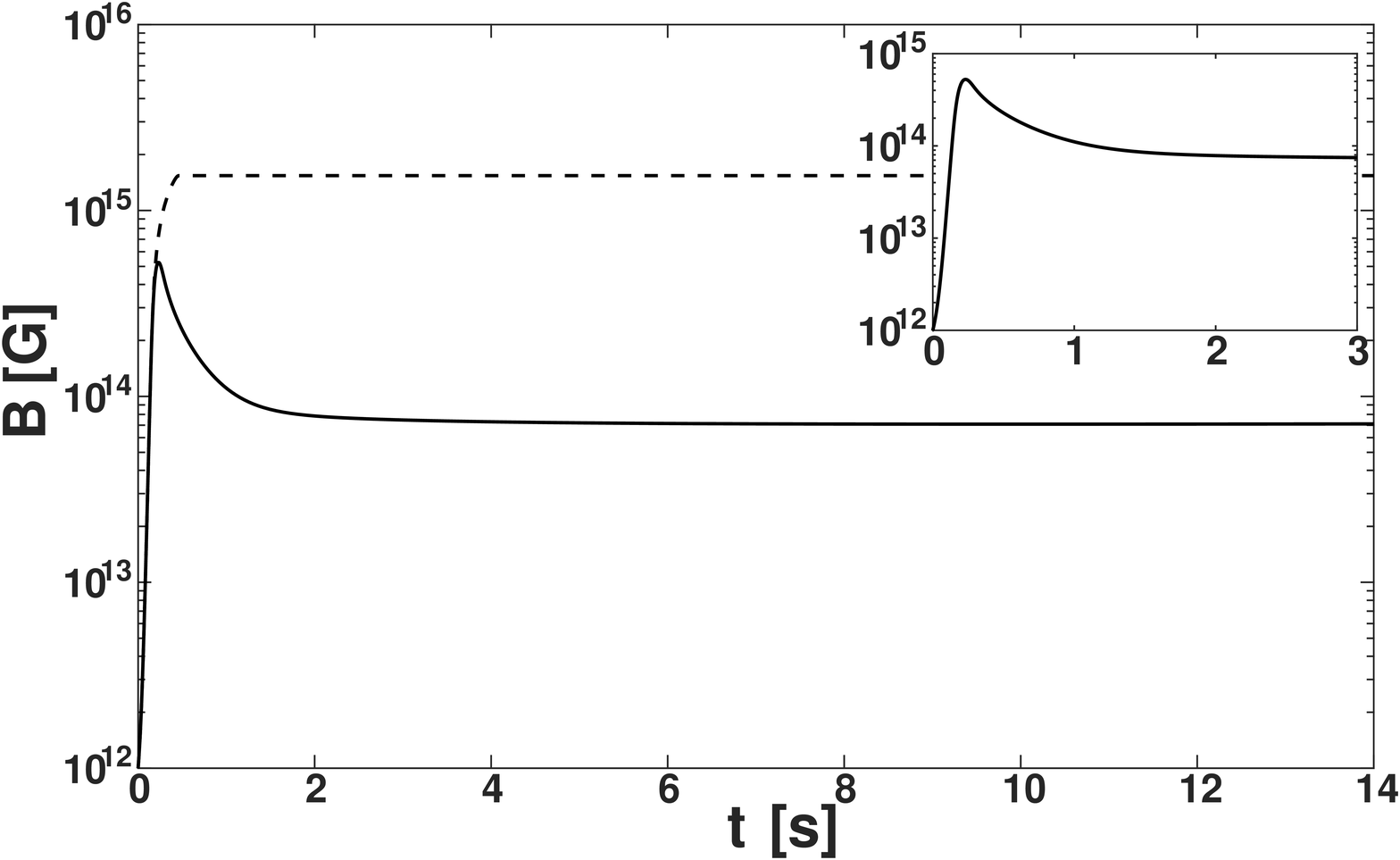}}
  \\
  \subfigure[]
  {\label{1c}
  \includegraphics[scale=.11]{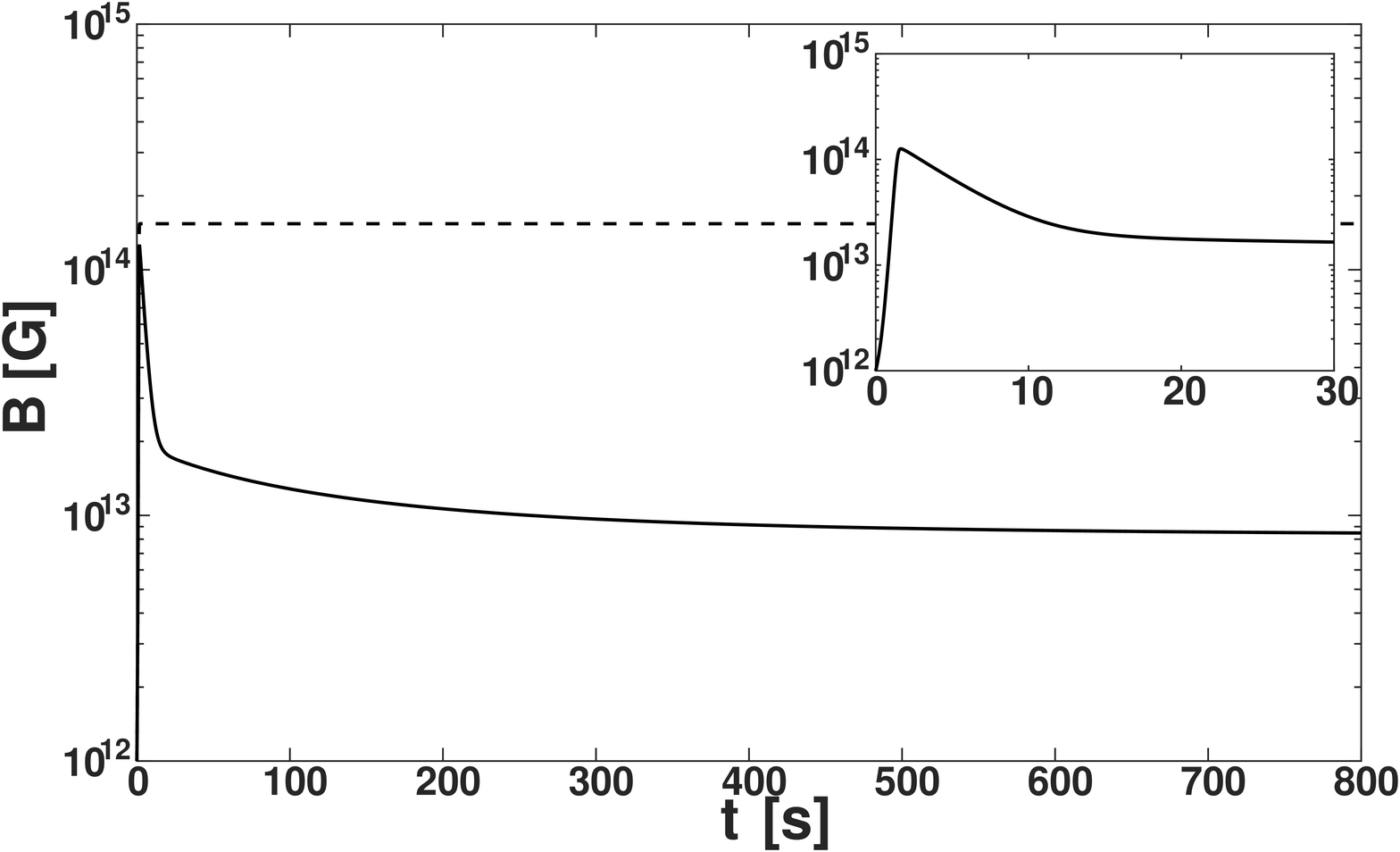}}
  \hskip-.7cm
  \subfigure[]
  {\label{1d}
  \includegraphics[scale=.11]{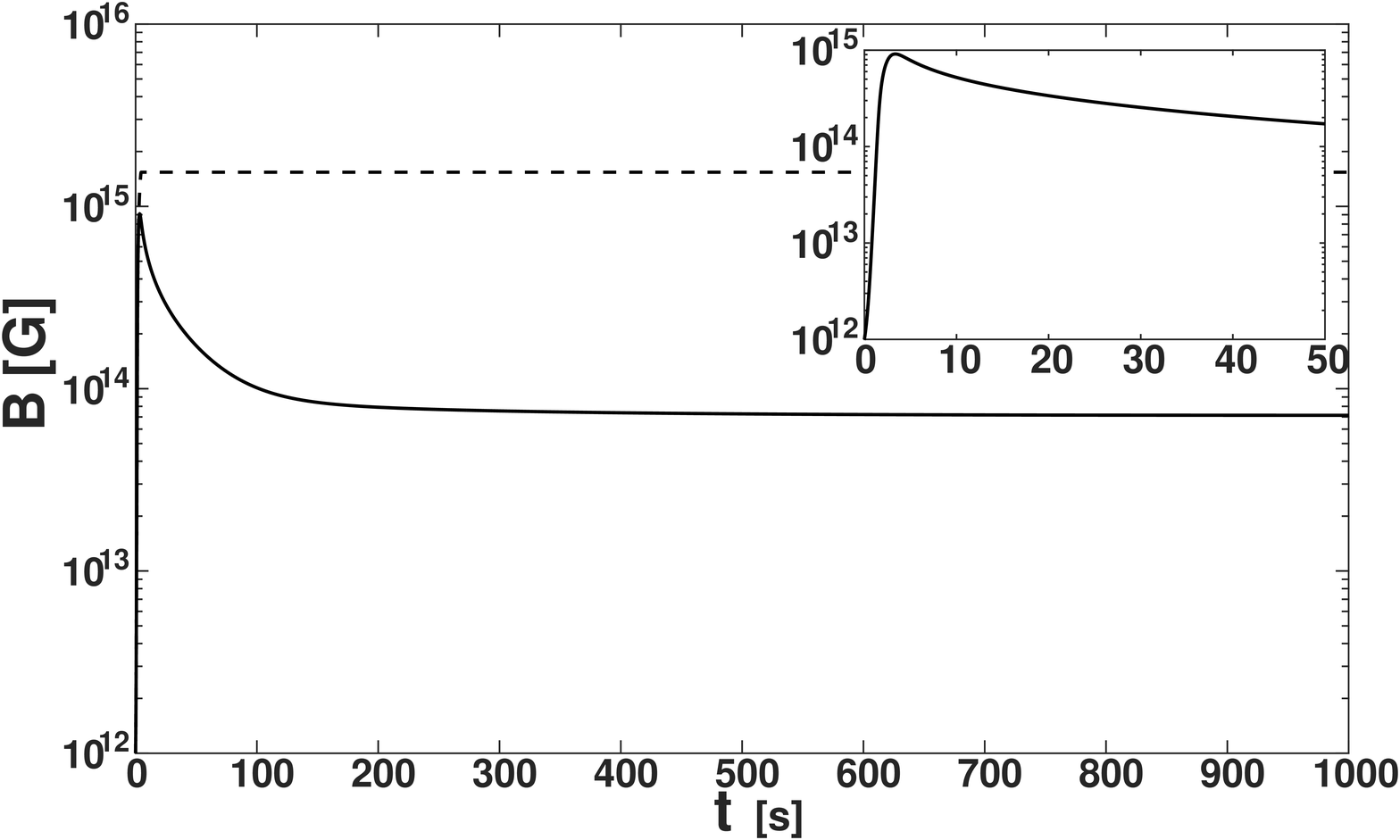}}
  \\
  \subfigure[]
  {\label{1e}
  \includegraphics[scale=.11]{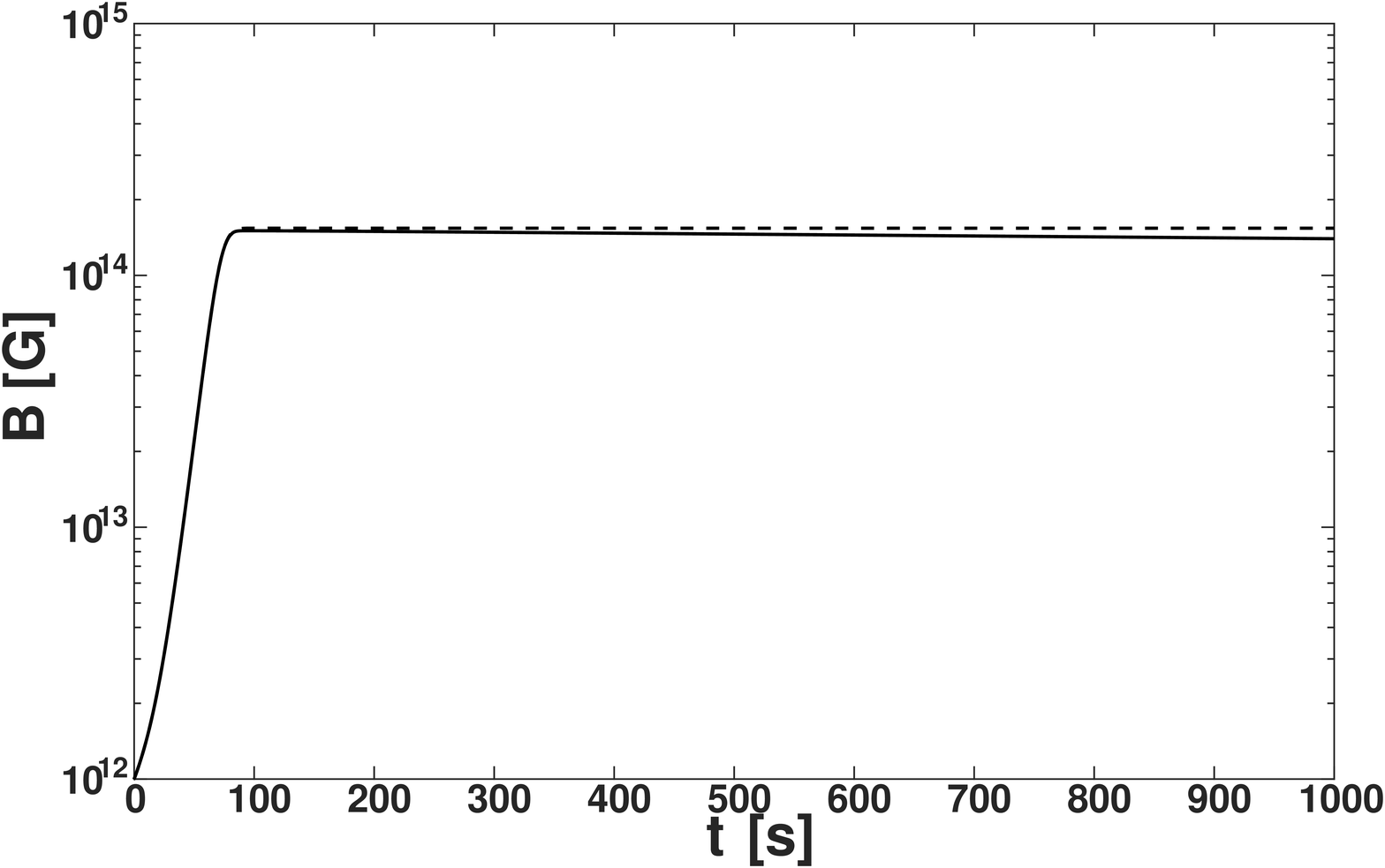}}
  \hskip-.7cm
  \subfigure[]
  {\label{1f}
  \includegraphics[scale=.11]{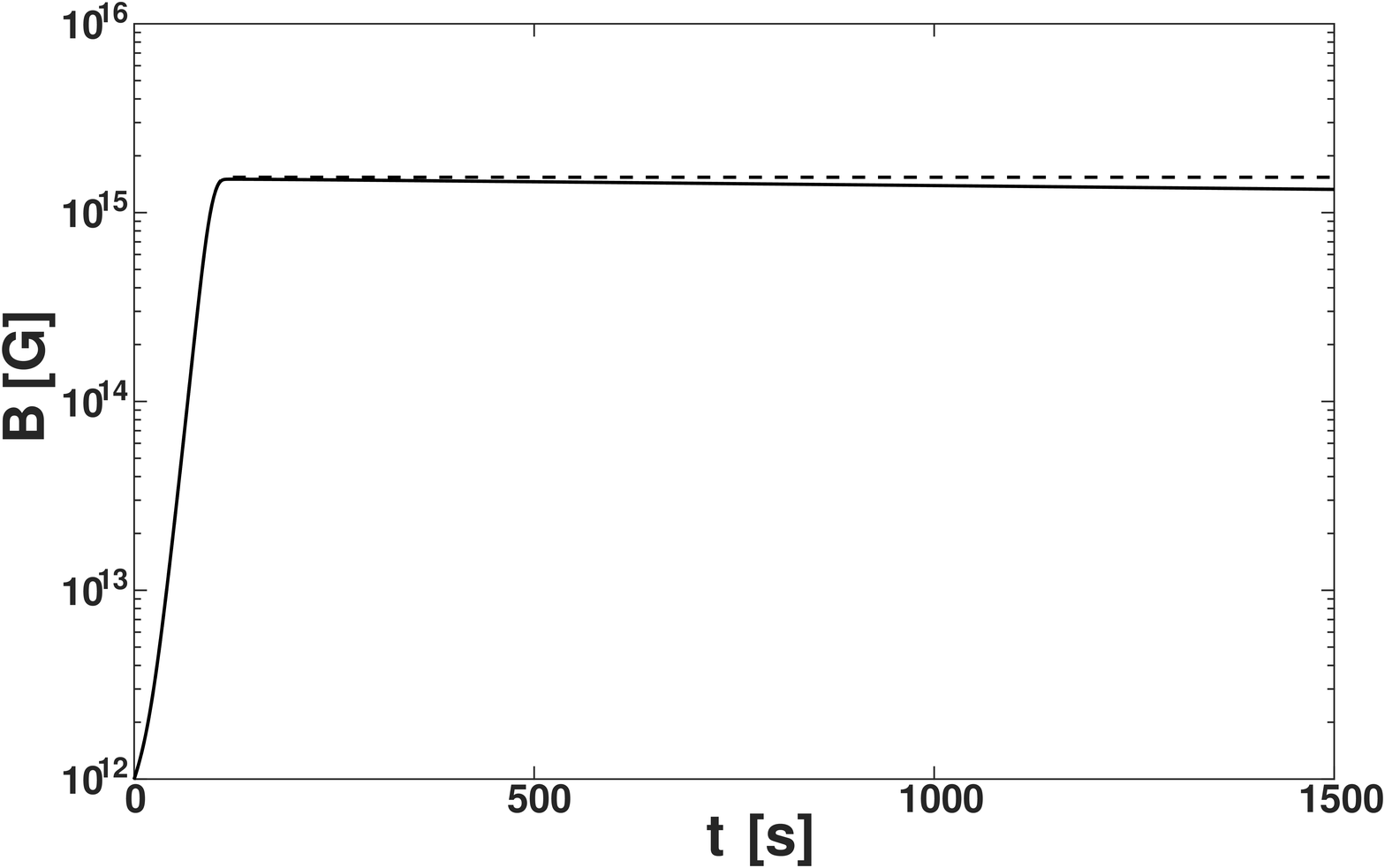}}
  \protect
  \caption{The evolution of the magnetic field for different $T_{0}$ and 
  $\Lambda_{\mathrm{max}}=1/k_{\mathrm{min}}$
  in HS when only $u$ and $d$ quarks are present.
  The panels~(a), (c), and~(e) correspond to $T_{0}=10^{8}\,\text{K}$,
  whereas the panels~(b), (d), and~(f) to $T_{0}=10^{9}\,\text{K}$.
  The panels~(a) and~(b) show the evolution of the field with
  $1\,\text{cm}<\Lambda<10\,\text{cm}$,
  the panels~(c) and~(d) with
  $10\,\text{cm}<\Lambda<10^{2}\,\text{cm}$,
  and the panels~(e) and~(f) with
  $5\times10^{2}\,\text{cm}<\Lambda<5\times10^{3}\,\text{cm}$.
  The insets in the panels (a)-(d) demonstrate the magnetic field behavior
  at small evolution times.
  Dashed lines represent the magnetic field evolution with the corresponding 
  initial conditions at the absence of the turbulence, i.e. when $K_d = 0$.
  \label{fig:Bfield}}
\end{figure}

We start with the case of quark matter in the core of HS, consisting only of $u$ and $d$ quarks, with $n_d = 2 n_u$. In Fig.~\ref{fig:Bfield}, we show the time evolution of the magnetic
field in this quark matter based on the numerical solution of Eqs.~(\ref{eq:syskindmlsH})-\eqref{eq:syskindmlsMd}
with the initial conditions chosen in Sec.~\ref{sec:MODEL}. The
numerical simulations were made for various initial temperatures and
minimal length scales. In all cases we start with the same magnetic
field $B_{0}=10^{12}\,\text{G}$.

One can see in Figs.~\ref{1a}-\ref{1d} that
the initial magnetic field experiences the exponential growth driven by
the electroweak interaction between quarks since $V_{5q}\neq0$. After
the magnetic field reaches the maximal value, it drops down about
one order of magnitude and then the field demonstrates a slow decrease. The appearance
of a peak in the profile of the time evolution of the magnetic field is owing to the MHD turbulence
effects accounted for in Eqs.~\eqref{eq:dhdt}-\eqref{eq:dmuddt} and~(\ref{eq:syskindmlsH})-\eqref{eq:syskindmlsMd}.

Let us define the scale of the magnetic field as $\Lambda=1/k$. Then
the effect of the turbulent motion of matter is more sizable for small scale magnetic
fields corresponding to a great $k_{\mathrm{min}}$, as seen in Eq.~(\ref{eq:Kd}).
To highlight this feature, in Fig.~\ref{fig:Bfield}, we plot the
evolution of the magnetic field at the absence of the turbulence,
i.e. when $K_{d}=0$, shown as dashed lines. In Figs.~\ref{1a}
and~\ref{1b}, for $1\,\text{cm}<\Lambda<10\,\text{cm}$,
as well as Figs.~\ref{1c} and~\ref{1d},
for $10\,\text{cm}<\Lambda<10^{2}\,\text{cm}$, the difference between
the turbulent and nonturbulent cases is clearly seen. For the greater
scale magnetic field, shown in Figs.~\ref{1e} and~\ref{1f},
corresponding to $5\times10^{2}\,\text{cm}<\Lambda<5\times10^{3}\,\text{cm}$,
the evolution of the magnetic field in the turbulent and nonturbulent
situations is less distinguishable in a limited time interval.

Then let us turn to the situation when both $u$, $d$, and $s$ quarks are present in a quark matter, with $n_u = n_d = n_s$. Such a quark matter may well exist in QS. As in Fig.~\ref{fig:Bfield}, here we also start with $B_0 = 10^{12}\,\text{G}$ and numerically solve Eqs.~(\ref{eq:syskindmlsH})-\eqref{eq:syskindmlsMd}. However we should take into account that the parameters in these equations are slightly different from those used to plot Fig.~\ref{fig:Bfield}. The discrepancy of parameters is discussed at the end of Sec.~\ref{sec:MODEL} and in Appendix~\ref{sec:TAUDCALC}.

\begin{figure}
  \centering
  \subfigure[]
  {\label{2a}
  \includegraphics[scale=.11]{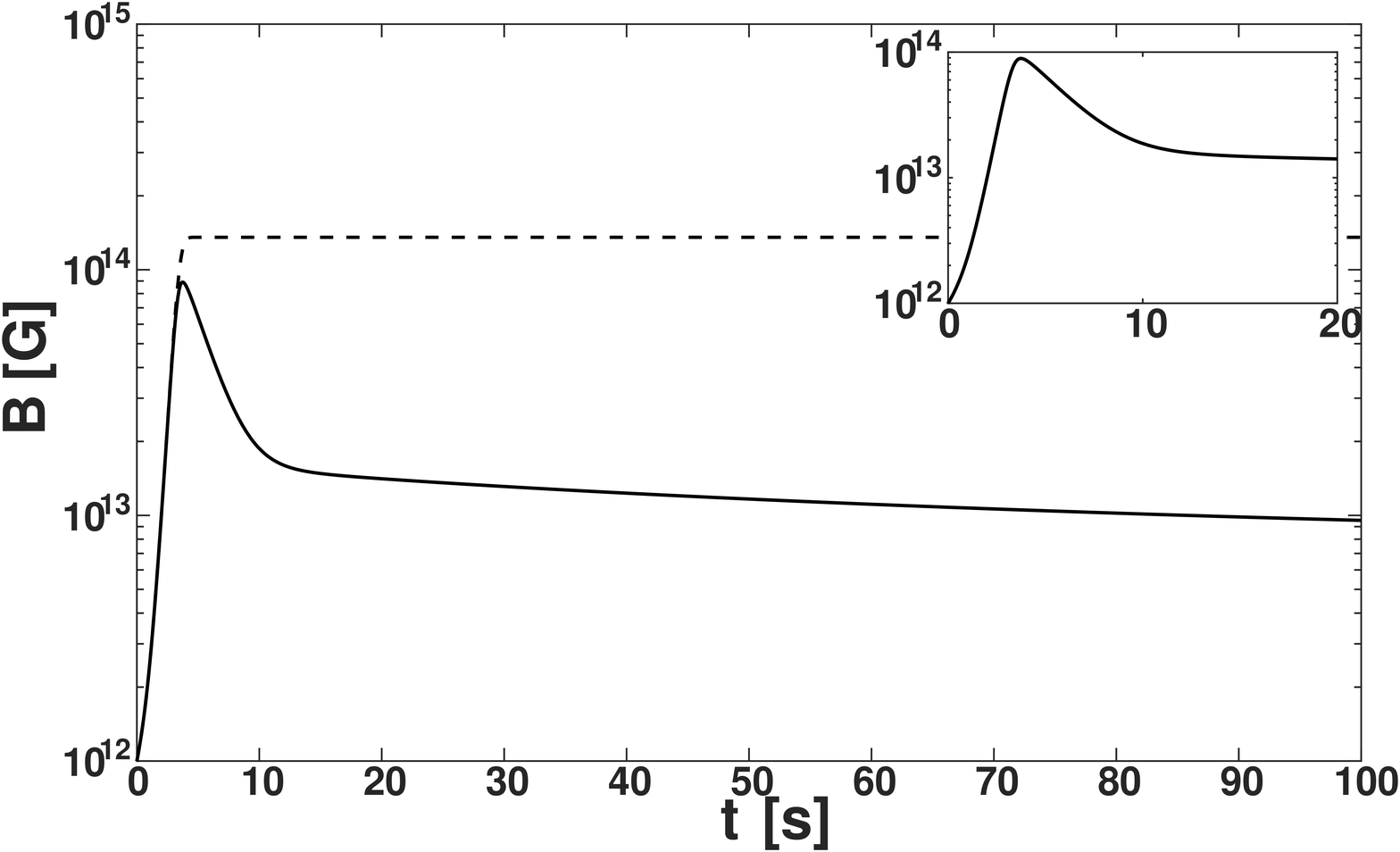}}
  \hskip-.7cm
  \subfigure[]
  {\label{2b}
  \includegraphics[scale=.11]{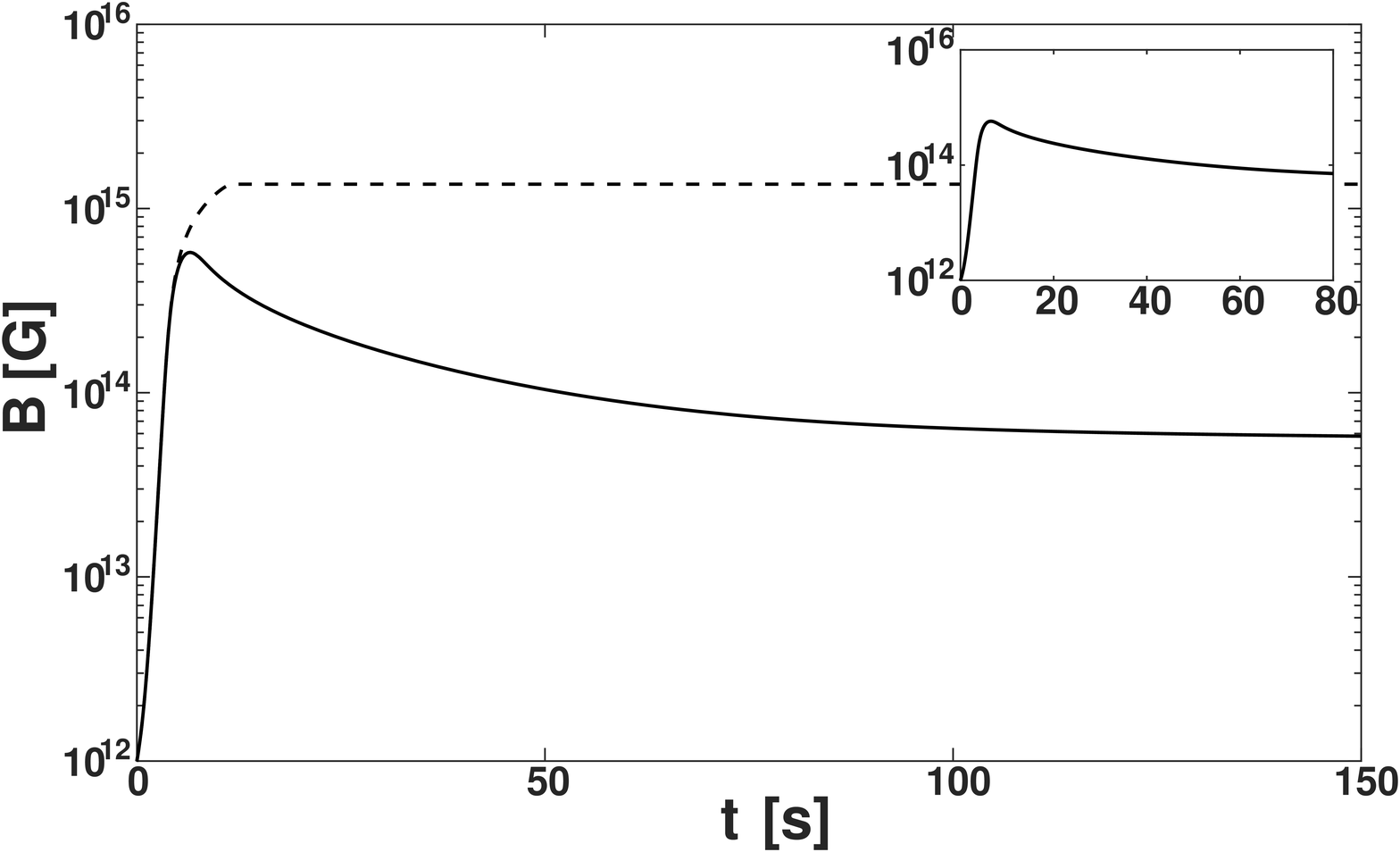}}
  \\
  \subfigure[]
  {\label{2c}
  \includegraphics[scale=.11]{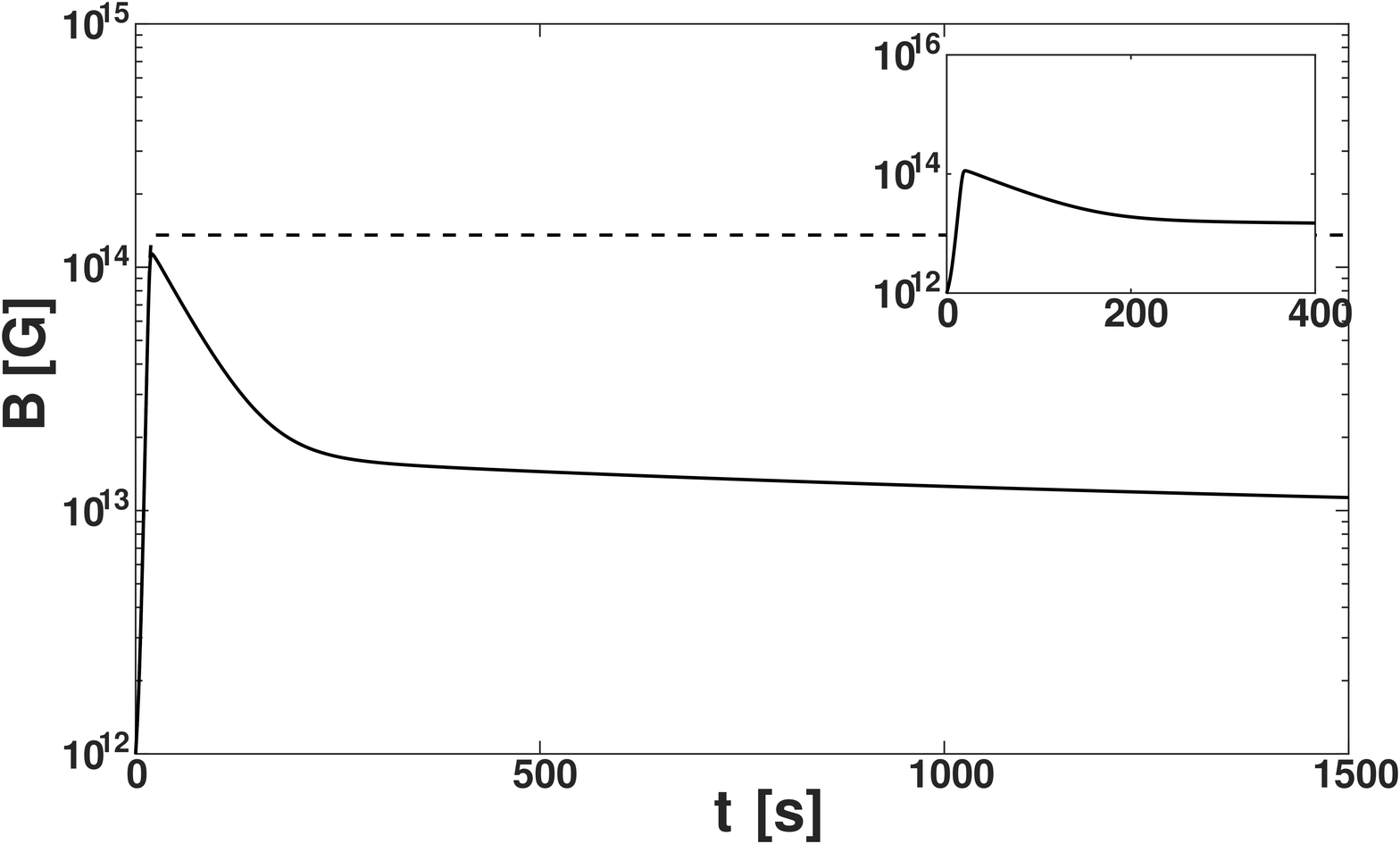}}
  \hskip-.7cm
  \subfigure[]
  {\label{2d}
  \includegraphics[scale=.11]{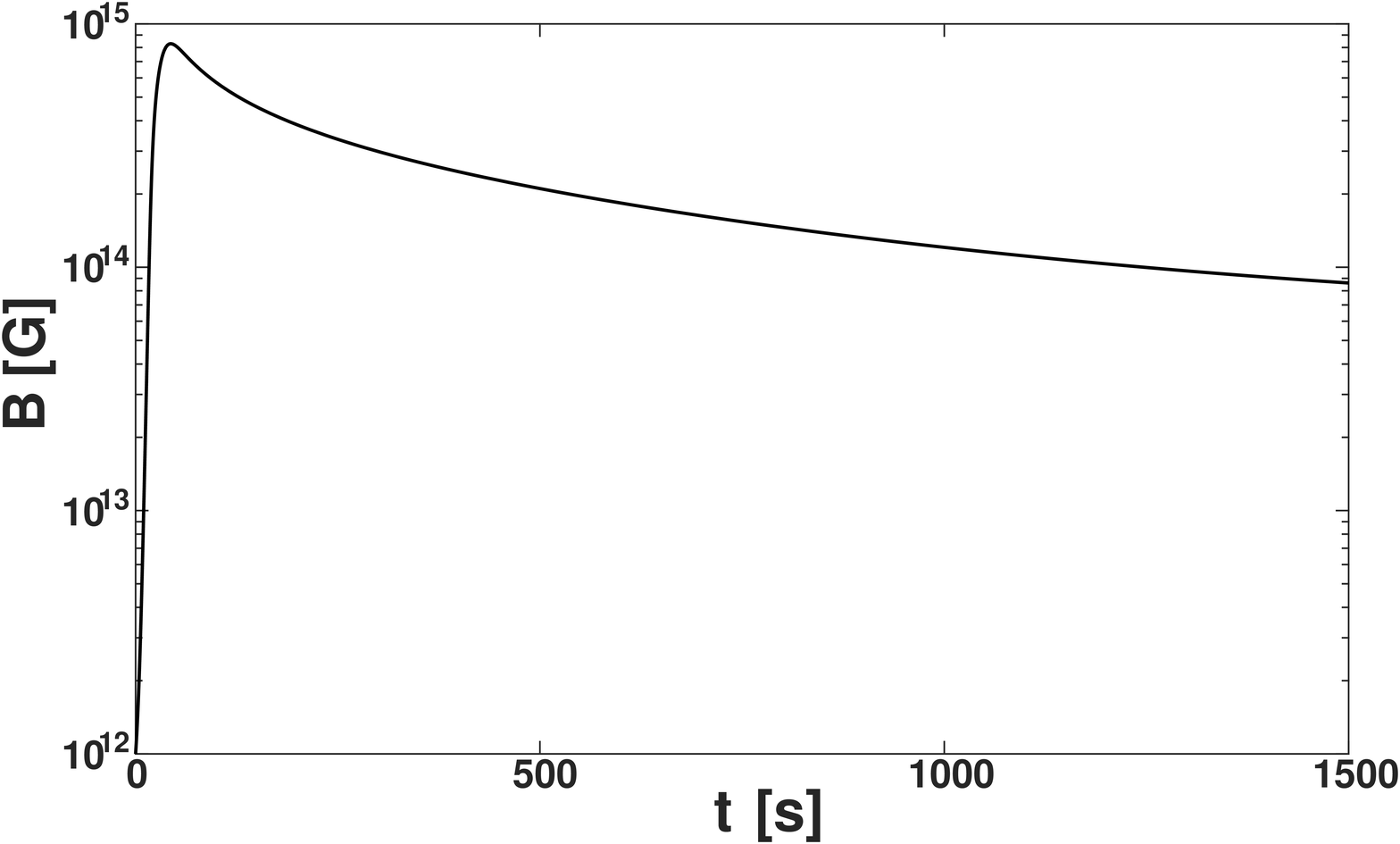}}
  \protect
  \caption{The evolution of the magnetic field for different $T_{0}$ and 
  $\Lambda_{\mathrm{max}}$
  in QS with $n_u = n_d = n_s$.
  The panels~(a) and~(c) correspond to $T_{0}=10^{8}\,\text{K}$,
  whereas the panels~(b) and~(d) to $T_{0}=10^{9}\,\text{K}$.
  The panels~(a) and~(b) show the evolution of the field with
  $10\,\text{cm}<\Lambda<10^2\,\text{cm}$, and
  the panels~(c) and~(d) with
  $50\,\text{cm}<\Lambda<5\times10^{2}\,\text{cm}$.
  The insets in the panels (a)-(c) demonstrate the magnetic field behavior
  at small evolution times.
  Dashed lines represent the magnetic field evolution with the corresponding 
  initial conditions at the absence of the turbulence, i.e. when $K_d = 0$.
  \label{fig:BfieldS}}
\end{figure}

The evolution of the magnetic field in this quark matter is shown in Fig.~\ref{fig:BfieldS}. The behavior of magnetic fields in this case qualitatively resembles that shown in Fig.~\ref{fig:Bfield}. However, due to the greater value of $K_d$ in the presence of $s$ quarks, the effect of the turbulence is more sizable. It means that one can study larger scale magnetic fields, which are still influenced by the matter turbulence (compare, e.g., Figs.~\ref{1c} and~\ref{2c}). In Fig.~\ref{2d} we do not show the evolution of the magnetic field in the non-turbulent matter---a dashed line is absent---because of the technical difficulties in the numerical solution of Eqs.~(\ref{eq:syskindmlsH})-\eqref{eq:syskindmlsMd} with the chosen coefficients.

As we will see in Sec.~\ref{sec:APPL}, the most interesting astrophysical applications correspond to the cases shown by solid lines in Figs.~\ref{1a}, \ref{1c}, \ref{2a}, and~\ref{2c}. In these situations, the maximal magnetic field $B_\mathrm{max}$ is about 30\% of the saturated value $B_\mathrm{sat}$: $B_\mathrm{max} \approx 0.3 B_\mathrm{sat}$. The saturated magnetic field, which corresponds to the horizontal dashed lines, is reached when there is an equipartition of the energy between the magnetic field and the energy reservoir: $B_\mathrm{sat}^2 \sim B_\mathrm{eq}^2$, where $B_\mathrm{eq}^2 = (\mu_u^2 + \mu_d^2) T_0^2$; cf. Eq.~\eqref{eq:Fqdef}. Hence, in the considered turbulent cases, the maximal energy of the growing magnetic field is only a small fraction of the background energy: $B_\mathrm{max}^2 \sim 0.1 B_\mathrm{eq}^2 \ll B_\mathrm{eq}^2$. Therefore $F_Q > 0$ in Eq.~\eqref{eq:Fqdef}.
%Thus, magnetic fields described in the present paper are really small fluctuations on a background.

In Sec.~\ref{sec:MODEL}, we assumed that the seed magnetic field is maximally helical. Such a magnetic field is force-free, i.e. $(\nabla \times \mathbf{B}) \sim \mathbf{B}$. One can check this fact using, e.g., the Chern-Simons wave, $\mathbf{B}(z,t) = B(t)(\mathbf{e}_x \sin kz + \mathbf{e}_y \cos kz)$, corresponding to the maximal positive helicity. Taking that $\mathbf{J} = (\nabla \times \mathbf{B})$ in Eq.~\eqref{eq:vFL}, we get the MHD turbulence is irrelevant for this initial field configuration since $\mathbf{v} \sim (\mathbf{J} \times \mathbf{B}) = 0$ in Eq.~\eqref{eq:vFL}. One can see in Figs.~\ref{fig:Bfield} and~\ref{fig:BfieldS} that the initial field evolution is not affected by the turbulence since solid and dashed lines almost coincide. Thus, at small evolution times, the exponential growth of the field is mainly driven by CME accompanied by the electroweak correction.

We have shown in Ref.~\cite{DvoSem16} that the MHD turbulence differently contributes the evolution of the magnetic helicity and the magnetic energy. Moreover, master Eqs.~\eqref{eq:drhodt}-\eqref{eq:dmuddt} are strongly nonlinear. Therefore, at certain moment of time $t=t_\mathrm{turb}$, the relation between the spectra corresponding to a maximally helical field is no longer valid, i.e. $h(k,t_\mathrm{turb}) \neq 2\rho_{\mathrm{B}}(k,t_\mathrm{turb})/k$, and hence $\mathbf{v} \neq 0$ in Eq.~\eqref{eq:vFL}. Thus, at $t > t_\mathrm{turb}$, the MHD turbulence can prevail the CME contribution. As one can see in Eqs.~\eqref{eq:dhdt}-\eqref{eq:dmuddt}, it should happen when the magnetic field reaches a certain strength since the turbulence contribution is quadratic in spectra compared to the CME contribution which is linear in spectra. It is this feature which can be observed in Figs.~\ref{fig:Bfield} and~\ref{fig:BfieldS}. Thus we have qualitatively described the behavior of the magnetic field when the MHD turbulence is accounted for.

\section{Application for the description of bursts of magnetars\label{sec:APPL}}

In this section, we discuss the implication of the results of the numerical
simulations in Sec.~\ref{sec:NUMRES} for the explanation of the
electromagnetic radiation of magnetars.

We mentioned already in Sec.~\ref{sec:INTR} that magnetars are mainly
detected by their electromagnetic radiation in the X-ray or soft $\gamma$-ray
regions as AXP or SGR. This electromagnetic activity is in the form
of bursts or flares, which range from short to giant, depending on
the duration and the total energy release. In all the events, a short
peak in the radiation is followed by a slowly decreasing tail. Peaks last $\sim0.1\,\text{s}$ in short bursts~\cite{Hup13} and $\sim(10-20)\,\text{s}$ in giant
flares~\cite{Fer01}. The tail of giant
flares can be up to $400\,\text{s}$~\cite{Fer01}. Some of the observational
data are summarized in Ref.~\cite{TurZanWat15}.

Nowadays there a lack in understanding of both the origin of the strong
magnetic fields in magnetars and the mechanism underlying their electromagnetic flashes. For example, in Ref.~\cite{Che96}, it was suggested
that flashes of SGR can be accounted for by the magnetized matter
emission through ruptures of the stellar crust in a sort of starquakes.
Presently, the electromagnetic activity of magnetars is believed to be associated with a twist of magnetic field lines in the stellar magnetosphere~\cite{ThoLyuKul02}. Such a twist should be triggered by a crustal motion. Ruptures and macroscopic slips of the crust are forbidden because of the huge pressure in the stellar matter. Moreover,
additional effective pressure is caused by a very strong magnetic field which permeates
the magnetar crust.

Recently the mechanism of plastic deformations in dense matter of a magnetar crust, driven by a thermoplastic wave (TPW), was proposed in Ref.~\cite{BelLev14}.
The TPW model, which turns out to be the most plausible
explanation of magnetar flashes now, was further developed in Ref.~\cite{Lan16}.
It was found in Ref.~\cite{Lan16} that the plastic flow becomes
irrelevant if $B\apprle10^{13}\,\text{G}$. However, the role of the
magnetic field in the core of a magnetar is unclear in this model.
The quantized vortices and fluxtubes in the superfluid stellar core
were suggested in Ref.~\cite{Lan16} to play an important role in
the TPW excitement.

In Ref.~\cite{Dvo16a}, we proposed a model for the magnetic field
generation in HS/QS based on the magnetic field instability driven
by the electroweak interaction of quarks. This model involves the
electroweak correction to CME. Note that the
existence of quark matter in a compact star is necessary since
CME in the presence of the electroweak interaction
can take place only if the chiral symmetry is unbroken~\cite{Dvo16c}.
Unlike a neutron star, there is a possibility to restore the chiral
symmetry in quark matter in HS/QS~\cite{BubCar16}. In Ref.~\cite{Dvo16a},
we described the amplification of the seed field $B_{0}=10^{12}\,\text{G}$, typical
for a pulsar, to $B=\left(10^{14}-10^{15}\right)\,\text{G}$, predicted
in magnetars~\cite{MerPonMel15,TurZanWat15}. The magnetic fields generated were
large scale with $10^{2}\,\text{m}<\Lambda<10\,\text{km}$.

Now we shall try to apply our model to explain the excitement of TPW. As we mentioned above, a physical mechanism triggering TPW is unclear. We can suggest that it is related to a fluctuation of a magnetic field, with the proper characteristics, in the inner crust or in the outer core of a star. We shall consider the generation of a fluctuation of the magnetic field in an outer core of a compact star where quark matter with unbroken chiral symmetry can be present. The time evolution of such a fluctuation and its characteristics, like the length scale and the magnetic field strength, will be studied. For this purpose we consider the evolution of smaller scale fields in our model accounting for the turbulence effects. We will reveal that the MHD turbulence is crucial to explain the observational data for magnetar bursts.

Let
us consider HS/QS in a thermal equilibrium with the initial temperature
$T_{0}=\left(10^{8}-10^{9}\right)\,\text{K}$. We shall take that
the quark matter density is $n_{0}=1.8\times10^{38}\,\text{cm}^{-3}$,
which may well exist in HS/QS. In this case, as results from the numerical
simulation in Sec.~\ref{sec:NUMRES}, one can generate the small
scale magnetic field with the time behavior resembling very much bursts and
flares of magnetars.

First, let us consider HS. The magnetic field evolution is shown in Fig.~\ref{fig:Bfield}. If we take $T_{0}=10^{8}\,\text{K}$, the initial pulse of
the magnetic field with $B\gtrsim10^{13}\,\text{G}$ lasts $\sim0.1\,\text{s}$,
see the inset in Fig.~\ref{1a}. We recall that the thermoplastic
flow is efficient if $B\gtrsim10^{13}\,\text{G}$~\cite{Lan16}.
As mentioned above, such a time interval is typical for a short burst
of a magnetar~\cite{Hup13}.

Changing the scale of the magnetic field, we can model giant flares of magnetars. Indeed, we can see in the inset in Fig.~\ref{1c}
that the duration of the initial pulse of the magnetic field is $\sim\left(10-15\right)\,\text{s}$,
which is quite close to that one observes in a giant flare of a magnetar~\cite{Fer01}.
The part of the slowly decreasing tail in which $B\gtrsim10^{13}\,\text{G}$
in Fig.~\ref{1c} lasts $\sim\left(300-400\right)\,\text{s}$,
which is again quite close to the total duration of a giant flare
of a magnetar~\cite{Fer01}. Even the shape of the initial pulse of
the magnetic field, obtained in our model, is similar to that observed
in a giant flare: it has a steeper leading edge and more slopping
tailing edge; cf. the inset in Fig.~\ref{1c} and Ref.~\cite{Fer01}. 

If we study the magnetic field evolution in QS, shown in Fig.~\ref{fig:BfieldS}, we can see that the scale of the magnetic field is slightly greater than that in Fig.~\ref{fig:Bfield}. It results from the fact that $K_d$ is greater in the quark matter with $n_u = n_d = n_s$. One can see that the results of numerical simulations depicted in Fig.~\ref{2a} can model an intermediate burst of a magnetar, with the duration of the initial pulse $\sim 10\,\text{s}$, described in Ref.~\cite{Ibr01}. Further enhancement of the scale of the magnetic field, compared to that in Figs.~\ref{2c} and~\ref{2d}, is inexpedient since the flares with the total duration longer than several hundred seconds are not observed.

In our model, the magnetic field is generated in a region
where the chiral symmetry is unbroken. It appears to happen close
to the stellar core. As mentioned above, the role of the magnetic
field in the stellar core is unclear in the TPW model~\cite{Lan16}. Now we can see that the generation of small scale fluctuations of the magnetic field in the stellar core can be among the physical processes which trigger TPW, since the time evolution of the magnetic field, shown in Figs.~\ref{fig:Bfield} and~\ref{fig:BfieldS}, resembles
very much the bursts and flares of magnetars. Note that, there are indications that magnetic fields can be important in plastic deformations of crystal lattice in rocks~\cite{Buc14}. This process is analogous to the TPW propagation.

According to our model for flashes of magnetars, the small scale fluctuation of the magnetic field, created near the core of a star, generates TPW which propagates through the magnetar crust towards the stellar surface. The speed of TPW was estimated in Ref.~\cite{BelLev14} to be $v_\mathrm{TPW} \sim (\chi \rho_m / \eta)^{1/2}$, where $\chi$ is the heat diffusion coefficient, $\rho_m$ is the magnetic energy density in the magnetar magnetosphere, and $\eta$ is the effective dynamic viscosity. The typical values of these parameters are~\cite{BelLev14} $\chi = 10\,\text{cm}^2\cdot\text{s}^{-1}$ and $\rho_m = 10^{28}\,\text{erg}\cdot\text{cm}^{-3}$, corresponding to $B_m = 4\times10^{14}\,\text{G}$. Basing on several assumptions, $\eta$ was estimated in Ref.~\cite{ChuYak05}. At temperatures $T \sim 10^8\,\text{K}$ and densities corresponding to the compact star crust, we can take that $\eta \sim 10^{12}\,\text{erg}\cdot\text{s}\cdot\text{cm}^{-3}$. It gives us $v_\mathrm{TPW} = 3.2\times10^8\,\text{cm}\cdot\text{s}^{-1}$. Assuming the crust depth $L_c = 1\,\text{km}$, we get that TPW will spend about  $\Delta t = L_c/v_\mathrm{TPW} = 3.2\times10^{-4}\,\text{s}$ to reach the magnetosphere. The obtained $\Delta t$ is much smaller even than the duration of a pulse $(0.2 - 0.3)\,\text{s}$ in Fig.~\ref{1a} corresponding to a short burst. Hence we can neglect the possible dispersion of TPW in our model.

Recently, in Ref.~\cite{LiLevBel16}, there was a suggestion that magnetar flares can be associated with small scale magnetic fields in the stellar core. For example, the initial magnetic field in the magnetar core necessary to induce a short wavelength Hall wave, which then can trigger TPW, was found in Ref.~\cite{LiLevBel16} to have $\sim 10\,\text{m}$ length scale, which is only one order of magnitude bigger than $\Lambda_\mathrm{max}$ used in Sec.~\ref{sec:NUMRES} to simulate a giant flare; cf. Fig.~\ref{1c}. Even greater value of $\Lambda_\mathrm{max} = 5\,\text{m}$ is used to plot Fig.~\ref{2c} corresponding to a flare in QS. Perhaps 3D simulations announced in Ref.~\cite{LiLevBel16} will reveal a closer coincidence with the results of our model.

It is interesting to mention that we can explain the flashes of magnetars,
ranging from short bursts to giant flares, by changing only one parameter in our model -- namely, the scale of the magnetic field.
The energetics of these events should depends on the wave length of
TPW excited. It is reasonable to assume that, if one generates
a magnetic field of a larger
scale, it will initiate TPW with a larger wave length. It is this feature which results from
our simulations: compare Figs.~\ref{1a} and~\ref{1c}
as well as Figs.~\ref{1b} and~\ref{1d}.

\section{Discussion\label{sec:CONCL}}

In the present work we have studied the evolution of strong magnetic
fields in quark matter driven by CME accompanied by the contribution of the electroweak interaction of quarks in Eq.~(\ref{eq:Jind}). To allow the existence
of CME we have assumed that the chiral symmetry
in quark matter is unbroken. Since magnetic fields were supposed to
be of a relatively small scale, we have accounted for the effects of
the MHD turbulence in quark matter by simulating the matter velocity with the Lorentz
force in Eq.~(\ref{eq:vFL}). Then we have derived the system of
the evolution equations for the spectra of the magnetic
helicity density and the magnetic energy density as well as for the
chiral imbalances of each type of quarks; cf.  Eqs.~(\ref{eq:syskindmlsH})-\eqref{eq:syskindmlsMd}. In Sec.~\ref{sec:NUMRES},
this system has been solved numerically with the initial conditions
corresponding to HS/QS.

In Sec.~\ref{sec:APPL}, we have analyzed the application of the
numerically obtained behavior of the magnetic fields for the interpretation
of the electromagnetic radiation of AXP and SGR. It has been revealed
that, under certain initial conditions, the time evolution of the
magnetic field obtained in our model resembles various flashes
in magnetars spanning from short bursts to giant flares. Thus small
scale magnetic fields generated in the stellar core could trigger
the electromagnetic processes in the magnetosphere of a compact star,
resulting in the bursts and flares, through, e.g., the mechanism of TPW propagating in the crust of a star, which was
recently proposed in Refs.~\cite{BelLev14,Lan16}. Hence the mechanism
of the magnetic field generation driven by the parity violating electroweak
interaction between fermions can be implemented in these compact stars.

%The instability of the magnetic field will dominate the matter turbulence if $\tau_\mathrm{D} \ll t_\mathrm{inst}$, where $t_\mathrm{inst}$ is the typical time of the magnetic field growth. In our case, the exponential growth of the magnetic field is driven by the electroweak interaction between quarks. Hence $t_\mathrm{inst} \sim \sigma_\mathrm{cond} \Lambda / \alpha_\mathrm{em} V_5$, which can be obtained on the basis of the modified Faraday equation in Ref.~\cite{Dvo16a}. Taking that $\mu_0 \sim 3 \times 10^2\,\text{MeV}$, $\Lambda \sim 10\,\text{cm}$, and $V_5 \sim 10\,\text{eV}$, which were adopted in Sec.s~\ref{sec:NUMRES} and~\ref{sec:APPL}, as well as using Eqs.~\eqref{eq:sigmaT} and~\eqref{eq:tauDfin}, we get that the considered condition is also valid. Thus, at the initial time interval, the turbulent motion of matter does not influence the exponential enhancement of magnetic fields shown in Figs.~\ref{fig:Bfield} and~\ref{fig:BfieldS}. Moreover, one can check that, in the considered time interval, one has that $\tau_\mathrm{D} \ll \Lambda$, i.e. the mean free path of fermions is much shorter than the length scale of the magnetic field.

It should be noted that the generation of peaks in Figs.~\ref{fig:Bfield} and~\ref{fig:BfieldS} is owing to the MHD turbulence. This turbulence becomes sizable when the magnetic field approaches its maximal value since the contribution of the turbulence is quadratic in the spectra $h(k,t)$ and $\rho_\mathrm{B}(k,t)$ in Eqs.~\eqref{eq:dhdt}-\eqref{eq:dmuddt}. Then, accounting for the anticorrelation of the matter temperature and the magnetic field strength, mentioned in Sec.~\ref{sec:MODEL} (see also Ref.~\cite{Dvo16b}), and Eq.~\eqref{eq:tauDfin}, we get that the mean free path of quarks becomes comparable with the length scale of the magnetic field. Hence,  turbulence effects stop the growth of the magnetic field driven by the electroweak correction to CME. This feature is well known in the studies of the turbulence (see, e.g., Ref.~\cite{Dav15}). Therefore, the described process results in the formation of peaks in the magnetic field profiles, which can be seen also in Figs.~\ref{1a}-\ref{1c} and~\ref{fig:BfieldS}.

Note that, in the present work, we have revealed that the time evolution of small scale magnetic fields in frames of our model looks similarly to flashes of magnetars. The actual physics mechanism of the interaction of these magnetic field fluctuations with  TPW, resulting in the wave excitement, has not been studied. This problem will be considered in our future works.

Another suggestion to generate strong magnetic field in a magnetar, which is applicable mainly in a neutron star, was put forward in Ref.~\cite{ThoDun93}. It is based on the MHD dynamo amplification of a seed field in the turbulent plasma of a protoneutron star. The matter turbulence is created by the dense flux of neutrinos emitted in a supernova explosion. The maximal magnetic field can be obtained from the equipartition of the energies of the magnetic field and the turbulent motion. The careful estimate made in Ref.~\cite{Spr09} gives one that the magnetic field with $B \sim 10^{14}\,\text{G}$ can be generated. In principle, such a magnetic field strength is enough to produce a magnetar burst (see the estimates in Sec.~\ref{sec:APPL}). However it is created in a protoneutron star when the neutrino flux is great, i.e. in the millisecond time interval after the supernova explosion. A magnetar burst, in its turn, is observed in a compact star at much later stages of its evolution~\cite{MerPonMel15}, when a star is thermally relaxed. Thus, our mechanism, which describes the generation of a strong magnetic field in degenerate quark matter, looks more plausible for the explanation of magnetar bursts.

%The influence of the turbulent motion of plasma on the generation of strong magnetic fields in protoneutron stars was studied recently in Refs.~\cite{Yam16a,Yam16b}. In these works, magnetic fields were created owing to the combined action of CME and the chiral vortical effect~\cite{SonSur09}. The action of the turbulence was accounted for by the direct analysis of the relativistic Navier-Stokes equation completed by the terms responsible for the chiral effects. In the present work we used an alternative approach based on the replacement of the plasma velocity with the Lorentz force in Eq.~\eqref{eq:vFL}. It allowed us to avoid the consideration of the nonlinear Navier-Stokes equation.

%\acknowledgments

\section*{Acknowledgments}

I am thankful to S.B.~Popov, A.I.~Rez, and R.~Turolla for useful communications, to V.B.~Se\-mi\-koz and D.D.Sokoloff for helpful discussions,
as well as to the Tomsk State University Competitiveness Improvement
Program and RFBR (research project No.~15-02-00293) for a partial support.

\appendix

\section{Calculation of drag time in degenerate quark matter\label{sec:TAUDCALC}}

In this appendix we shall calculate the drag time $\tau_\mathrm{D}$, introduced in Sec.~\ref{sec:MODEL}, due to Coulomb collisions
of quarks.

In plasma consisting of $u$ and $d$ quarks, there are three types
of collisions: $ud\to ud$, $uu\to uu$, and $dd\to dd$. Let us,
first, consider the $ud\to ud$ reaction. The matrix element has the
form,
\begin{equation}\label{eq:Mud}
  \mathcal{M}_{ud} =
  \frac{\mathrm{i}e_{u}e_{d}}{t}
  \bar{u}(p_{1}')\gamma^{\mu}u(p_{1}) \cdot
  \bar{d}(p_{2}')\gamma_{\mu}d(p_{2}),
\end{equation}
where $t=(p_{1}'-p_{1})^{2}$ is the Mandelstam variable, $u(p)$
and $d(p)$ are the bispinors of $u$ and $d$ quarks corresponding
to the initial $p_{1,2}$ and final $p_{1,2}'$ momenta, and $\gamma^{\mu}=\left(\gamma^{0},\bm{\gamma}\right)$
are the Dirac matrices. The matrix element squared takes the form,
\begin{equation}\label{eq:Mud2}
  |\mathcal{M}_{ud}|^{2} = \frac{e_{u}^{2}e_{d}^{2}}{t^{2}}
  \mathrm{tr}
  \left(
    \rho_{2}'\gamma_{\mu}\rho_{2}\gamma_{\nu}
  \right) \cdot
  \mathrm{tr}
  \left(
    \rho_{1}'\gamma^{\mu}\rho_{1}\gamma^{\nu}
  \right),
\end{equation}
where
\begin{equation}\label{eq:densmatr}
  \rho_{1,2}=\frac{1}{2}
  \gamma_{\mu}p_{1,2}^{\mu}
  \left(
    1-\chi_{1,2}\gamma^{5}
  \right),
  \quad
  \rho_{1,2}'=\frac{1}{2}\gamma_{\mu}p_{1,2}^{\prime\mu},
\end{equation}
are the spin density matrices for the initial and final quark states~\cite[p.~111]{BerLifPit82}.
Here $\chi=\pm1$ is the helicity of the initial quarks ($\chi=+1$
for right quarks and $\chi=-1$ for left particles) and $\gamma^{5}=\mathrm{i}\gamma^{0}\gamma^{1}\gamma^{2}\gamma^{3}$.
Equation~(\ref{eq:densmatr}) implies the assumption of unpolarized final
states. Using Eq.~(\ref{eq:densmatr}), one can transform Eq.~(\ref{eq:Mud2})
to the form,
\begin{equation}\label{eq:M2Mv}
  |\mathcal{M}_{ud}|^{2} =
  \frac{e_{u}^{2}e_{d}^{2}}{2t^{2}}
  \left[
    s^{2}+u^{2}+\chi_{1}\chi_{2}
    \left(
      s^{2}-u^{2}
    \right)
  \right],
\end{equation}
where $s=\left(p_{1}+p_{2}\right)^{2}$ and $u=\left(p_{1}-p_{2}'\right)^{2}$
are the Mandelstam variables.

The total probability of the $ud$ scattering per unit time reads
\begin{equation}\label{eq:Wpm}
  W_{ud} =2
  \left(
    W_{+}+W_{-}
  \right),
  \quad
  W_{+} = \frac{Ve_{u}^{2}e_{d}^{2}}{4(2\pi)^{8}}
  \left\langle
    \frac{s^{2}}{t^{2}}
  \right\rangle,
  \quad
  W_{-}=\frac{Ve_{u}^{2}e_{d}^{2}}{4(2\pi)^{8}}
  \left\langle
    \frac{u^{2}}{t^{2}}
  \right\rangle,
\end{equation}
where $V$ is the normalization volume and the averaging of a variable
$X=X(p_{1},\dotsc,p_{2}')$ is defined as
\begin{align}\label{eq:averdef}
  \left\langle
    X
  \right\rangle = &   
  \int\frac{\mathrm{d}^{3}p_{1}\mathrm{d}^{3}p_{2}
  \mathrm{d}^{3}p_{1}'\mathrm{d}^{3}p_{2}'}
  {E_{1}E_{2}E_{1}'E_{2}'}
  X%(p_{1},\dotsc,p_{2}')
  \delta^{4}(p_{1}+p_{2}-p_{1}'-p_{2}')
  \nonumber
  \\
  & \times
  f(E_{1}-\mu_{u})
  \left[
    1-f(E_{1}'-\mu_{u})
  \right]
  f(E_{2}-\mu_{d})
  \left[
    1-f(E_{2}'-\mu_{d})
  \right].
\end{align}
Here $E=|\mathbf{p}|$ is the energy of a massless quark, $f(E)=\left[\exp(\beta E)+1\right]^{-1}$
is the Fermi-Dirac distribution function, $\beta=1/T$ is the reciprocal
temperature, and $\mu_{u,d}$ are the chemical potentials of $u$
and $d$ quarks. Note that $W_{\pm}$ in Eq.~(\ref{eq:Wpm}) correspond
to $\chi_{1}\chi_{2}=\pm1$ in Eq.~(\ref{eq:M2Mv}). In Eq.~(\ref{eq:M2Mv}),
we take into account all possible initial helicities and sum over
the polarization states of the outgoing quarks.

\begin{figure}
  \centering
  \includegraphics[scale=0.2]{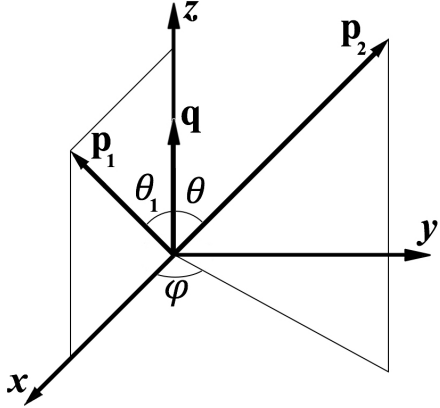}
  \protect
  \caption{The relative position of the momenta for the integration
  over $\mathbf{p}_{1}$ and $\mathbf{p}_{2}$.
  \label{fig:momint}}
\end{figure}

Let us first compute $\left\langle s^{2}/t^{2}\right\rangle $. After
the integration over $\mathbf{p}_{2}'$ with help of the momentum
conservation delta function, Eq.~(\ref{eq:averdef}) takes the form,
\begin{align}\label{eq:s2t2p2}
  \left\langle
    \frac{s^{2}}{t^{2}}
  \right\rangle = & 
  \int
  \frac{\mathrm{d}^{3}p_{1}\mathrm{d}^{3}p_{1}'\mathrm{d}^{3}p_{2}}
  {E_{1}E_{2}E_{1}'E_{2}'}
  \frac{
  \left[
    E_{1}E_{2} -
    \left(
      \mathbf{p}_{1} \cdot \mathbf{p}_{2}
    \right)
  \right]^{2}}
  {\left[
    E_{1}E_{1}' -
    \left(
      \mathbf{p}_{1} \cdot \mathbf{p}_{1}'
    \right)
  \right]^{2}}
  \delta(E_{1}+E_{2}-E_{1}'-E_{2}')
  \nonumber
  \\
  & \times
  f(E_{1}-\mu_{u})
  \left[
    1-f(E_{1}'-\mu_{u})
  \right]
  f(E_{2}-\mu_{d})
  \left[
    1-f(E_{2}'-\mu_{d})
  \right],
\end{align}
where $E_{2}'=|\mathbf{q}-\mathbf{p}_{2}|$ and $\mathbf{q}=\mathbf{p}_{1}'-\mathbf{p}_{1}$.
Then we represent $\mathrm{d}^{3}p_{2}=E_{2}^{2}\mathrm{d}E_{2}\mathrm{d}\cos\theta\mathrm{d}\varphi$
in Eq.~(\ref{eq:s2t2p2}) and choose the relative position of the momenta as shown in Fig.~\ref{fig:momint}.
In this case, $\left( \mathbf{p}_{1}\cdot\mathbf{p}_{2} \right) = E_{1}E_{2} ( \cos\theta\cos\theta_{1} + \sin\theta\sin\theta_{1}\cos\varphi )$
and
\begin{equation}
  \delta(E_{1}+E_{2}-E_{1}'-E_{2}') =
  \frac{E_{2}'}{qE_{2}}
  \delta(\cos\theta-\cos\theta_{0}),
  \quad
  \cos\theta_{0} \approx \cos\theta_{1} +
  \frac{q}{2E_{2}}\sin^{2}\theta_{1}
  \left(
    1+\frac{E_{2}}{E_{1}}
  \right).
\end{equation}
Moreover, taking into account that the quark matter is highly degenerate,
one gets that $f(E_{1,2}-\mu_{u,d})\left[1-f(E_{1,2}'-\mu_{u,d})\right]\approx T\delta(E_{1,2}-\mu_{u,d})$.
Finally, after the integration over $\mathbf{p}_{2}'$, Eq.~(\ref{eq:s2t2p2})
is transformed to
\begin{align}\label{eq:s2t2q}
  \left\langle
    \frac{s^{2}}{t^{2}}
  \right\rangle = &
  6\pi T^{2}\mu_{d}^{2}
  \int\mathrm{d}^{3}p_{1}\mathrm{d}^{3}p_{1}'
  \frac{\sin^{4}\theta_{1}\delta(E_{1}-\mu_{u})}
  {q
  \left[
    E_{1}E_{1}' -
    \left(
      \mathbf{p}_{1}\cdot\mathbf{p}_{1}'
    \right)
  \right]^{2}}.
\end{align}
Equation~(\ref{eq:s2t2q}) can be further simplified if we change the
integration variable $\mathbf{p}_{1}'\to\mathbf{q}=\mathbf{p}_{1}'-\mathbf{p}_{1}$
and notice that $E_{1}E_{1}'-\left(\mathbf{p}_{1}\cdot\mathbf{p}_{1}'\right)\approx(q^{2}/2)\sin^{2}\theta_{1}$.
Eventually we obtain that
\begin{equation}\label{eq:s2t2fin}
  \left\langle
    \frac{s^{2}}{t^{2}}
  \right\rangle =
  192\pi^{3}T^{2}\mu_{d}^{2}\mu_{u}^{2}
  \int_{0}^{\infty}
  \frac{\mathrm{d}q}{\left(q^{2}+\omega_{p}^{2}\right)^{3/2}} =
  192\pi^{3}\mu_{d}^{2}\mu_{u}^{2}\frac{T^{2}}{\omega_{p}^{2}},
\end{equation}
where we introduce the plasma frequency in the degenerate $ud$ matter~\cite{Dvo16a,BraSeg93},
\begin{equation}\label{eq:omegap}
  \omega_{p} = \frac{1}{\sqrt{3}\pi}
  \sqrt{e_{u}^{2}\mu_{u}^{2}+e_{d}^{2}\mu_{d}^{2}},
\end{equation}
to avoid the infrared divergence in Eq.~(\ref{eq:s2t2fin}).

Analogously one can compute $W_{-}$. The direct calculation shows
that $W_{-}=W_{+}$. It gives one the total probability of the $ud$
scattering in the form,
\begin{equation}\label{eq:Wud}
  W_{ud} = \frac{9V}{4\pi^{3}}
  \frac{e_{u}^{2}e_{d}^{2}\mu_{d}^{2}\mu_{u}^{2}}
  {e_{u}^{2}\mu_{u}^{2}+e_{d}^{2}\mu_{d}^{2}},
\end{equation}
where we use Eqs.~(\ref{eq:Wpm}), (\ref{eq:s2t2fin}), and~(\ref{eq:omegap}).

Now let us consider the scattering of identical quarks, e.g., the $uu\to uu$
reaction. The matrix element of this process is
\begin{equation}\label{eq:Muu}
  \mathcal{M}_{uu} = \mathrm{i}e_{u}^{2}
  \left[
    \frac{1}{t}
    \bar{u}(p_{1}')\gamma^{\mu}u(p_{1}) \cdot
    \bar{u}(p_{2}')\gamma_{\mu}u(p_{2}) -
    \frac{1}{u}
    \bar{u}(p_{2}')\gamma^{\mu}u(p_{1}) \cdot
    \bar{u}(p_{1}')\gamma_{\mu}u(p_{2})
  \right].
\end{equation}
The matrix element in Eq.~(\ref{eq:Muu}) squared takes the form~\cite[p.~327]{BerLifPit82},
\begin{equation}\label{eq:Muu2}
  |\mathcal{M}_{uu}|^{2}=\frac{e_{u}^{4}}{2}
  \left[
    \frac{s^{2}+u^{2}}{t^{2}} +
    \frac{s^{2}+t^{2}}{u^{2}}+\chi_{1}\chi_{2}
    \left(
      \frac{s^{2}-u^{2}}{t^{2}}+\frac{s^{2}-t^{2}}{u^{2}}
    \right)
  \right].
\end{equation}
Performing similar calculations as in the $ud$ case and using Eq.~(\ref{eq:Muu2}),
one obtains the total probability of the $uu\to uu$ reaction as
\begin{equation}\label{eq:Wuu}
  W_{uu} =
  \frac{9V}{2\pi^{3}}
  \frac{e_{u}^{4}\mu_{u}^{4}}{e_{u}^{2}\mu_{u}^{2}+e_{d}^{2}\mu_{d}^{2}}
  \left\{
    1+\frac{2}{3}\frac{\omega_{p}^{2}}{\mu_{u}^{2}}
    \left[
      \ln
      \left(
        \frac{\mu_{u}}{\sqrt{2}\omega_{p}}
      \right)+1
    \right]
  \right\}.
\end{equation}
The expression for the total probability of the $dd\to dd$ scattering
$W_{dd}$ can be obtained on the basis of Eq.~(\ref{eq:Wuu}) by
replacing $e_{u}\to e_{d}$ and $\mu_{u}\to\mu_{d}$.

Let us define the drag time as
\begin{equation}
  \tau_{\mathrm{D}} =
  \frac{N_{u}}{W_{uu}}+\frac{N_{d}}{W_{dd}}+\frac{\sqrt{N_{u}N_{d}}}{W_{ud}},
\end{equation}
where $N_{u,d}=n_{u,d}V$ are the total densities of $u$ and $d$
quarks and $n_{u,d}$ are the corresponding number densities. Using
Eqs.~(\ref{eq:Wud}) and~(\ref{eq:Wuu}) we obtain the following
expression for $\tau_{\mathrm{D}}$:
\begin{equation}
  \tau_{\mathrm{D}} = \frac{4\pi}{27T^{2}}
  \left(
    e_{u}^{2}\mu_{u}^{2}+e_{d}^{2}\mu_{d}^{2}
  \right)
  \left(
    \frac{2}{e_{u}^{4}\mu_{u}}+\frac{2}{e_{d}^{4}\mu_{d}} +
    \frac{2}{e_{u}^{2}e_{d}^{2}\sqrt{\mu_{u}\mu_{d}}}
  \right),
\end{equation}
where we keep only the leading term.
%and take into account that $\mu_{u,d}=\left(3\pi^{2}n_{u,d}\right)^{1/3}$.
Finally, basing on the values of $e_q$ and $\mu_q$ in Sec.~\ref{sec:MODEL}, one gets for $\tau_{\mathrm{D}}$
\begin{equation}\label{eq:tauDfin}
  \tau_{\mathrm{D}} = 2.5\frac{\mu_{0}}{\alpha_{\mathrm{em}}T^{2}},
\end{equation}
which is used in Sec.~\ref{sec:MODEL}. If one studies QS, the value of $\tau_{\mathrm{D}}$ will be about 1.2 times greater than that in Eq.~\eqref{eq:tauDfin}. It should be noted that
the dependence of $\tau_{\mathrm{D}}$ in Eq.~(\ref{eq:tauDfin})
on $\alpha_{\mathrm{em}}$ and $T$ coincides with that found in Ref.~\cite{Kel73}, where collisions between electrons and protons in degenerate matter were studied.

\end{document}